\documentclass[aps,pre,groupedaddress,showkeys,showpacs,amsmath,superscriptaddress,nofootinbib]{revtex4-1}
\usepackage{amssymb}
\usepackage{amsmath}
\usepackage{graphicx}
\usepackage{amsbsy}
\usepackage{ulem}
\usepackage{bm}
\usepackage{color}
\usepackage{gensymb}
\usepackage{caption}
\usepackage{subcaption}
\begin{document}

\title{Universal effects of solvent species on the stabilized structure of a protein}

\author{Tomohiko Hayashi}
\affiliation{Institute of Advanced Energy, Kyoto University, Uji, Kyoto 611-0011, Japan}

\author{Masao Inoue}
\affiliation{Institute of Advanced Energy, Kyoto University, Uji, Kyoto 611-0011, Japan}
\affiliation{Graduate School of Science, Chiba University, 1-33 Yayoi-cho, Inage, Chiba 263-8522, Japan}
\affiliation{Molecular Chirality Research Center, Chiba University, 1-33 Yayoi-cho, Inage, Chiba 263-8522, Japan}

\author{Emanuele Petretto} 
\affiliation{Dipartimento di Scienze Molecolari e Nanosistemi, 
Universit\`{a} Ca' Foscari di Venezia
Campus Scientifico, Edificio Alfa,
via Torino 155,30170 Venezia Mestre, Italy}

\author{Tatjana \v{S}krbi\'{c}} 
\affiliation{Dipartimento di Scienze Molecolari e Nanosistemi, 
Universit\`{a} Ca' Foscari di Venezia
Campus Scientifico, Edificio Alfa,
via Torino 155,30170 Venezia Mestre, Italy}

\author{Achille Giacometti} 
\email{achille.giacometti@unive.it}
\affiliation{Dipartimento di Scienze Molecolari e Nanosistemi, 
Universit\`{a} Ca' Foscari di Venezia
Campus Scientifico, Edificio Alfa,
via Torino 155,30170 Venezia Mestre, Italy}

\author{Masahiro Kinoshita}
\thanks{Corresponding author}
\email{kinoshit@iae.kyoto-u.ac.jp}
\affiliation{Institute of Advanced Energy, Kyoto University, Uji, Kyoto 611-0011, Japan}

\date{\today}

\begin{abstract}
We investigate the effects of solvent specificities on the stability of the native structure (NS) of a protein on the basis of our free-energy function (FEF). We use CPB-bromodomain (CBP-BD) and apoplastocyanin (apoPC) as representatives of the protein universe and water, methanol, ethanol, and cyclohexane as solvents. The NSs of CBP-BD and apoPC consist of 66$\%$ $\alpha$-helices and of 35$\%$ $\beta$-sheets and 4$\%$ $\alpha$-helices, respectively. In order to assess the structural stability of a given protein immersed in each solvent, we contrast the FEF of its NS against that of a number of artificially created, misfolded decoys possessing the same amino-acid sequence but significantly different topology and a-helix and $\beta$-sheet contents. In the FEF, we compute the solvation entropy using the morphometric approach combined with the integral equation theories, and the change in electrostatic (ES) energy upon the folding is obtained by an explicit atomistic but simplified calculation. The ES energy change is represented by the break of protein-solvent hydrogen bonds (HBs), formation of protein intramolecular HBs, and recovery of solvent-solvent HBs. Protein-solvent and solvent-solvent HBs are absent in cyclohexane. We are thus able to separately evaluate the contributions to the structural stability from the entropic and energetic components. We find that for both CBP-BD and apoPC, the energetic component dominates in methanol, ethanol, and cyclohexane, with the most stable structures in these solvents sharing the same characteristics described as an association of a-helices. In particular, those in the two alcohols are identical. In water, the entropic component is as strong as or even stronger than the energetic one, with a large gain of translational, configurational entropy of water becoming crucially important, so that the relative contents of $\alpha$-helix and $\beta$-sheet and the content of total secondary structures are carefully selected to achieve sufficiently close packing of side chains. If the energetic component is excluded for a protein in water, the priority is given to closest side-chain packing, giving rise to the formation of a structure with very low $\alpha$-helix and $\beta$-sheet contents. Our analysis, which requires minimal computational effort, can be applied to any protein immersed in any solvent and provides robust predictions that are quite consistent with the experimental observations for proteins in different solvent environments, thus paving the way toward a more detailed understanding of the folding process.
\end{abstract}
\maketitle
\section{Introduction}
\label{sec:introduction}
Protein folding is one of the most fundamental examples of biological self-assembly processes, and unveiling its mechanism is a crucially important task for understanding life phenomena. Shortly after it was established by Anfinsen \cite{Anfinsen73} that the primary sequence encoded all the necessary information to obtain the three-dimensional native fold, it also became clear that protein folding could be achieved only in aqueous environment. The hydrophobic effect has long been discussed as an essential factor driving a protein to fold. \cite{Ball08} According to the conventional view \cite{Kauzmann59,Tanford62}, when a nonpolar group, which cannot participate in hydrogen bonding of water, comes in contact with water, structuring of water occurs near the nonpolar group for retaining as many hydrogen bonds (HBs) as possible, causing entropic instability. The amount of such unstable water is reduced upon the burial of nonpolar groups within the protein interior. This is the only origin of the hydrophobic effect. We have proposed a substantially different view: \cite{Kinoshita08,Yoshidome12,Kinoshita13,Oshima15} Protein folding in aqueous solution under physiological conditions leads to a gain of the translational, configurational entropy of water in the entire system ($\Delta S_{S}>0$); and this is the hydrophobic effect. The reduction of water crowding (i.e., entropic correlation among water molecules coexisting with the protein) is the principal contributor to the water-entropy gain.\cite{Yoshidome12,Kinoshita13,Oshima15} The folding is also accompanied by a decrease in the protein intramolecular interaction energy ($\Delta E_{PP}<0$), increase in the protein-water interaction energy ($\Delta E_{PS}>0$), decrease in the water-water interaction energy ($\Delta E_{SS}<0$)  due to the structural reorganization of water released to the bulk, and loss of the protein conformational entropy ($\Delta S_{P}<0$). Each of $\Delta E_{PP}$, $\Delta E_{PS}$, $\Delta E_{SS}$ consists of electrostatic (ES) and van der Waals (vdW) components. It is clear that $\Delta S_{S}$, $\Delta E_{PP}$, and $\Delta E_{SS}$ favorably promote the folding whereas $\Delta E_{PS}$ and $\Delta S_{S}$ oppose it. (More detailed information is presented in Table \ref{tab:tab1}.) However, the task of accounting for each of these physical factors in a theoretical model is daunting and has not been accomplished yet.
\begin{table}[htbp]
  \caption{ Physical factors promoting or opposing protein folding in aqueous solution under physiological conditions. SS=water entropy (translational, configurational entropy is the largest contributor), EPP=protein intramolecular interaction energy, EPS=protein-water interaction energy, ESS=water-water interaction energy, and SP=protein conformational entropy. $\Delta X$ denotes the change in $X$ upon protein folding. The subscript “P”, “S”, “ES”, and “vdW” signify “protein”, “solvent”, “electrostatic” and “van der Waals”, respectively. $E_{PP,ES}$, for example, denotes the electrostatic component of $E_{PP}$. $\Delta E_{PS,ES}+\Delta E_{SS,ES}>0$. It is apparent that $\Delta E_{PP}+\Delta E_{SS}+\Delta E_{PS} -T (\Delta S_{S}+\Delta S_{P}) <0$  ($T$ is the absolute temperature).}
  \label{tab:tab1}
  \begin{tabular}{lcc}
    \hline \hline
    & \text{Entropic} & \text{Energetic} \\
    \hline
    \text{Promoting}  & $\Delta S_S>0$ & $\Delta E_{PP}=\Delta E_{PP,ES}+\Delta E_{PP,vdW}; \Delta E_{PP,vdW}<0, \Delta E_{PP,vdW}<0 $\\
    \text{Promoting}  & $\Delta S_S>0$ & $\Delta E_{SS}=\Delta E_{SS,ES}+\Delta E_{SS,vdW}; \Delta E_{SS,vdW}<0, \Delta E_{SS,vdW}<0 $\\
    \text{Opposing}   & $\Delta S_P>0$ & $\Delta E_{PS}=\Delta E_{PS,ES}+\Delta E_{PS,vdW}; \Delta E_{PS,vdW}>0, \Delta E_{PS,vdW}>0 $\\
    \hline \hline
  \end{tabular}
\end{table}
A clue to the protein folding mechanism and relative magnitudes of the physical factors mentioned above is to know how the stability of the native structure (NS) of a protein is affected when water is replaced by a different, less polar solvent. A previous study \cite{Pace04} suggested that the NS would become unstable in most of less polar solvents such as ethanol and methanol, very stable in nonpolar solvents such as cyclohexane, and even more stable in vacuum. Experimental observations \cite{Hirota98} indicate that alcohol induces a protein to form α-helices and the resultant helical structure is independent of the alcohol species. The aim of this study is to elucidate the effects of solvent specificities on the structural stability of a protein and deepen the understanding of the folding process. In the preceding work,\cite{Hayashi17} we studied the structures stabilized in water and a hard-sphere solvent whose particle diameter and packing fraction were set at those of water. We chose a particular protein, protein G, consisting of 27$\%$ α-helices and 39$\%$ β-sheets. The NS and a number of artificially created, misfolded decoys were examined, and the structure giving lowest value to our free-energy function (FEF)11−13 was identified in water or the hard-sphere solvent. In this study, on the other hand, we resort to two alternative and wildly different (from each other and from protein G) proteins: one including only α-helices and the other only $\beta$-sheets (strictly, it consists of 35$\%$ $\beta$-sheets and 4$\%$ a-helices). Furthermore, we consider a total of four different solvents matching those exploited in the aforementioned experiments: water, methanol, ethanol, and cyclohexane, thus changing the polarity (see Table \ref{tab:tab2}), molecular size, and packing fraction. The proteins in vacuum are also considered. As in our preceding work,11 the NS and a number of decoys are examined for each protein and the same FEF is applied to the identification of the most stable fold in each solvent.
In Table \ref{tab:tab1}, when a different solvent is considered, “water” should be replaced by “solvent”. We use a simplified and yet realistic representation of all of the entropic and energetic contributions to thermodynamics of protein folding where the FEF fully accounts for $\Delta S_{S}>0$, neglects$\Delta S_{S}0$ because only compact structures are considered, assumes that $\Delta E_{PS,vdw}>0$, $\Delta E_{SS,vdw}<0$, $\Delta E_{PP,vdw}<0$   cancel out, and represents  $\Delta E_{PS,ES}>0$, $\Delta E_{SS,ES}<0$, and $\Delta E_{PP,ES}<0$  in terms of the change in the sum of protein-solvent, solvent-solvent, and protein-protein hydrogen bonding energies. The relative magnitudes of the physical factors are dependent on the solvent species. In cyclohexane modeled as a completely nonpolar solvent, $\Delta E_{PS,ES}=0$ and $\Delta E_{SS,ES}=0$. In vacuum,$\Delta S_{S}=0$, $\Delta E_{PS,vdw}=0$,$\Delta E_{SS,vdw}=0$, $\Delta E_{PS,ES}=0$,$\Delta E_{SS,ES}=0$, but $\vert \Delta E_{PP,vdw}\vert \ll \vert \Delta E_{PP,ES}\vert$ and $\Delta E_{PP,vdw}$ can be neglected. The solvation entropy, which is denoted simply by $S$ hereafter, is dependent on the solvent species: It becomes smaller as the molecular diameter $d_S$ increases and/or the packing fraction $\eta_S$decreases. $S$ is also influenced by the solvent-solvent interaction potential. The effect of $\Delta E_{PP,ES}<0$  relative to that of $\Delta E_{PS,ES} +\Delta E_{SS,ES}>0$  becomes larger as the molecular polarity of the solvent decreases. $S$ is computed using the morphometric approach (MA)\cite{Yoshidome12,Oshima15,Koning04,Roth06} combined with the integral equation theories. \cite{Kinoshita08,Hansen06,Kusalik88a,Kusalik88b,Kinoshita96,Cann97}  The change in the sum of hydrogen bonding energies upon the folding is obtained by an explicit atomistic but simplified calculation. For each model solvent, we estimate the parameters used in calculating the FEF by referring to the experimental data. As a great advantage, the evaluation of the FEF is accomplished in $\approx $ 1 s per protein structure on a standard workstation, so that considerably many different structures can be examined quite efficiently at once.\cite{Hayashi17,Yoshidome09,Yasuda11}
The results common to both the $\alpha$-helix-rich and $\beta$-sheet-rich proteins can briefly be summarized as follows. The NS is identified as the most stable structure in water, manifesting the reliability of our FEF. The structures stabilized in methanol, ethanol, cyclohexane, and vacuum are characterized by high $\alpha$-helix contents (associated a-helices). Those in methanol and ethanol are identical. The structure stabilized in cyclohexane is also quite similar. In vacuum, the structure possessing the maximum number of protein intramolecular hydrogen bonds (IHBs) is the most stable. It is an association of $\alpha$-helices as structural units but the number of the units is smaller than that in methanol, ethanol, and cyclohexane. It is worthwhile to add that the NS of the $\alpha$-helix-rich protein is somewhat different from the structures stabilized in the three other solvents and vacuum. These results are suggestive that only in water a variety of structures with different $\alpha$-helix and $\beta$-sheet contents are simultaneously stabilized depending on the amino-acid sequence. The physical reasons for the effects of solvent species are discussed in detail. It is also argued that the results obtained are consistent with the aforementioned experimental observations for proteins in different solvent environments. This study shows that our FEF can be applied to any protein immersed in any solvent, underpins the stabilization effects of different solvents on the folding mechanism within a single theoretical framework, and highlights the unique role played by water.

\section{Model and Theory}
\begin{figure}[htpb]
  \centering
    \includegraphics[width=.8\linewidth]{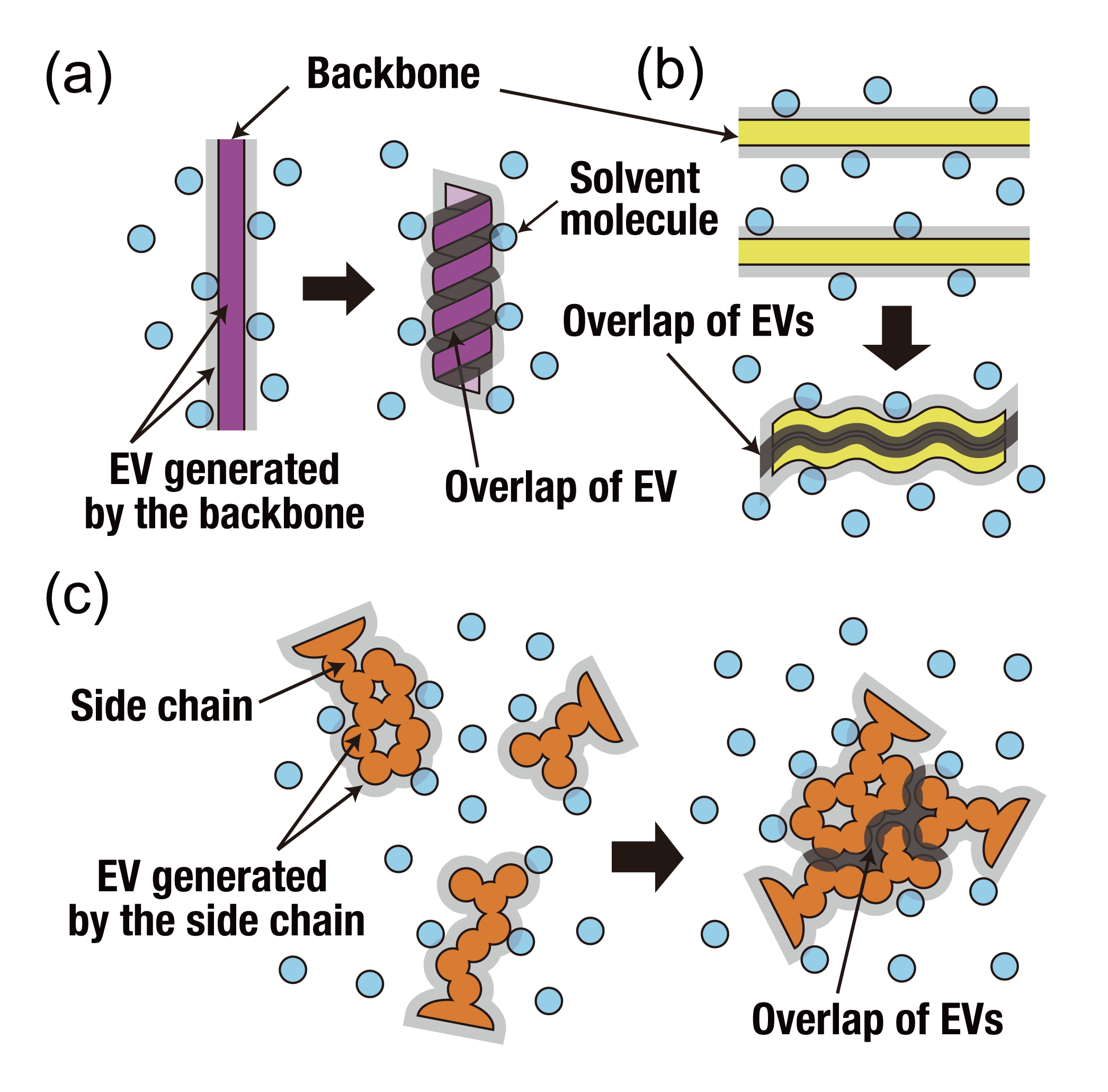}
  \caption{(a) Formation of $\alpha$-helix by a portion of the backbone. (b) Formation of $\beta$-sheet by portions of the backbone. (c) Close packing of side chains. Overlap of an excluded volume (EV) or EVs occurs. Consequently, the total volume available to the translational displacement of solvent molecules in the entire system increases by the overlapped volume, leading to a gain of solvent entropy.
  \label{fig:fig1}}
\end{figure}

\subsection{Entropic excluded-volume effect}
The solvent plays a significant role in protein folding through the entropic excluded-volume (EV) effect \cite{Yoshidome12,Kinoshita13,Oshima15} as pictorially illustrated in Fig.\ref{fig:fig1}. The EV is defined as the volume of the space unaccessible to the centers of solvent molecules. The formation of secondary structures ($\alpha$-helix and $\beta$-sheet) by the backbone and the packing of side chains lead to the reduction of the total EV. This reduction results in a gain of the solvent entropy.\cite{Kinoshita13,Oshima15,Harano05,Yasuda10,Yasuda12} The close packing of side chains with a variety of geometric features is especially important from the entropic point of view. We note that there are both translational and rotational contributions to the solvent-entropy gain, but the former is much larger than the latter even when the solvent is water. \cite{Kinoshita08,Yoshidome08} As detailed below, we exploit a molecular model for the solvent because the EV effect cannot be described by the dielectric continuum model.
   The solvent molecules are entropically correlated because the presence of each solvent molecule generates an EV for the other solvent molecules. We refer to this solvent-solvent entropic correlation as “solvent crowding”. Upon protein folding, the solvent crowding is significantly reduced, and this protein-solvent many-body correlation makes an essential contribution to the solvent-entropy gain. It also plays a pivotal role in cold and pressure denaturation of a protein.\cite{Yoshidome12,Kinoshita13,Oshima15} The many-body correlation is not taken into account in the Asakura-Oosawa (AO) theory. \cite{Asakura54,Asakura58} See Appendix A for more information.
   In the conventional view of the hydrophobic effect,\cite{Kauzmann59,Tanford62} when water is replaced by a nonpolar solvent, the contact between the solvent and a nonpolar group is no more unstable, leading to the absence of solvophobic effect. In our view, by contrast, the solvophobic effect does exist due to the aforementioned solvent crowding.
   \subsection{Free-energy function}
   For a protein immersed in a solvent, the FEF ($F$) is defined as11−13
   \begin{eqnarray}
     \label{eq1}
     \frac{F}{k_BT_0} &=& \frac{\Lambda}{k_BT_0}-\frac{S}{k_B}
   \end{eqnarray}
   where $T$ is the absolute temperature and $k_B$  is the Boltzmann constant ($T$ is set at $T_0$). $F/(k_BT_0)$ and its energetic and entropic components, $\Lambda/(k_BT_0)$ and $−S/k_B$, respectively, are dependent on the protein structure. $\Delta F$ is defined as “$F$ for a decoy structure” minus “$F$ for the NS”:
   \begin{eqnarray}
     \label{eq2}
     \Delta F &=& F\text{(a decoy structure)} - F\text{(the NS)}.
   \end{eqnarray}
$\Delta \Lambda$  and −$\Delta S$ are also defined in similar fashions. When $\Delta F$ is negative, the decoy structure is more stable than the NS for the specified solvent.
\subsection{Solvent models}
\begin{table}[htbp]
  \caption{Dielectric constant $\epsilon$ of solvent (at 293 K) as a measure of its polarity.}
  \label{tab:tab2}
  \begin{tabular}{lcccc}
    \hline \hline
    \text{Solvent}& \text{Cyclohexane} & \text{Ethanol} & \text{Methanol} & \text{Water} \\
    \hline
    $\epsilon$(-)  &        2.02        &        25.3    &        33.0     &     80.4 \\
    \hline \hline
  \end{tabular}
\end{table}
\begin{table}[htbp]
  \caption{Summary of the solvent characteristic parameters used in this study. Here, “D” and “Q” denote “dipole” and “quadrupole”, respectively.}
  \label{tab:tab3}
  \begin{tabular}{lcccc}
    \hline \hline
    \text{Solvent}      & \text{Diameter} $d_s$(nm) & \text{Packing fraction} $\eta_s$(-) & \text{Multipoles} & \text{Reduced Dipole moment} $\mu_{S}^{*}$(-) \\
    \hline
    \text{Water}        &        0.280              &        0.383                        &        \text{D+Q} &     2.768 \\
    \text{Methanol}     &        0.347              &        0.326                        &        \text{D}   &     1.295 \\
    \text{Ethanol}      &        0.403              &        0.352                        &        \text{D}   &     1.029 \\
    \text{Cyclohexane}  &        0.560              &        0.512                        &        \text{-}   &     0 \\
    \hline \hline
  \end{tabular}
\end{table}
In order to span the entire spectrum of solvent polarity, we consider the four solvents, water, methanol, ethanol, and cyclohexane (see Table \ref{tab:tab2}) using the experimentally measured density at 298 K and 1 atm for each solvent. As in our earlier works,\cite{Hayashi17,Yoshidome12,Yasuda11} the model water molecule adopted is a hard sphere with diameter $d_S$=0.28 nm ($\rho_S d_S^3$=0.7317; $\rho_S$ is the number density of bulk solvent) in which a point dipole and a point quadrupole of tetrahedral symmetry are embedded. \cite{Kinoshita08,Kusalik88a,Kusalik88b}
   Methanol and ethanol molecules are modeled as dipolar hard spheres. Their diameters, dipole moments, and packing fractions are estimated as follows. Assuming that monohydric alcohol molecules interact through the Stockmayer potential, for a polar gas of a monohydric alcohol Monchick and Mason \cite{Monchick61} determined its potential parameters by adapting the Chapman-Enskog theory combined with experimental viscosity data. We modify the parameters in the dipole-dipole interaction part to make the potential pertinent to a liquid state. Specifically, the dipole moment $\mu_S$ is replaced by a larger, effective one accounting for the polarization effect,\cite{Jorgensen86} and the orientations of the dipole moments of two molecules are chosen so that the dipole-dipole attractive interaction can be maximized (i.e., the most probable orientations \cite{Kinoshita91} are chosen). The diameter $d_S$ is then evaluated as the distance between two molecular centers at which the potential equals $k_BT_0$ ($T_0$=298 K). This procedure is followed for both methanol and ethanol. Once $d_S$ is estimated, the packing fraction $\eta_S=\rho_S d_S^3/6$ can be calculated using the experimental data of methanol or ethanol density at 298 K and 1 atm. The results are $d_S$=0.347 nm, $\mu_{S}^{{*}^2}=\mu_{S}^2/(d_S^3 k_B T_0)$ =1.678, and $\eta_{S}$=0.326 for methanol and $d_S$=0.403 nm, $\mu_{S}^{{*}^2}$=1.059, and $\eta_{S}$=0.352 for ethanol. (For water, $\mu^{{*}^2}$=7.662 and $\eta_S$=0.383: The effect of molecular polarizability is incorporated in the water model using a self-consistent mean field theory. \cite{Kusalik88a,Kusalik88b})
A cyclohexane molecule is modeled as a neutral hard sphere with diameter $d_S$=0.56 nm. According to X-ray crystallography data at $186−280$K, cyclohexane is in a plastic solid state.30 The molecules are rotationally disordered about the lattice points of the face-centered cubic cell with a length of each side of 0.87 nm.30 The density calculated from the solid data is slightly higher than that of liquid cyclohexane. Hence, the effective diameter of molecules in liquid state $d_S$ is also slightly smaller than the value estimated from the solid data. We set $d_S$ at the double of the molecular diameter of water. Using the experimentally measured density of cyclohexane at 298 K and 1 atm, we obtain $\eta_S$=0.512. Table \ref{tab:tab3} summarizes the relevant quantities for each solvent.
   A word of caution is in order here. In our model systems, the pressure varies from solvent to solvent and departs from 1 atm. For example, the pressure of cyclohexane is much higher than 1 atm because the solvent-solvent vdW attractive interaction is neglected. However, the solvation entropy is determined by $\eta_S$  and $d_S$ rather than by the pressure, and these are carefully estimated as described above. In the preceding paper,\cite{Hayashi17} we considered water and a hard-sphere solvent sharing the same values of  $\eta_S$  and $d_S$. The pressure of the hard-sphere solvent is much higher than that of water, nevertheless the solvation entropies are not significantly different.
   \subsection{Calculation of entropic component}
   For a protein with a prescribed structure, the solvation entropy $S$ in Eq. (\ref{eq1}) is calculated under the isochoric condition. A thermodynamic quantity of solvation calculated under the isochoric condition is not influenced by the expansion or compression of bulk solvent upon the solute insertion.\cite{Cann97} Hence, it is physically more insightful than that calculated under the isobaric condition. The calculation of $S$ is performed by a hybrid method combining the integral equation theory (IET)\cite{Kinoshita08,Hansen06,Kusalik88a,Kusalik88b,Cann97} and the morphometric approach (MA). \cite{Yoshidome12,Oshima15,Koning04,Roth06}  An angle-dependent version \cite{Kinoshita08,Kusalik88a,Kusalik88b,Cann97} of the IET is employed for water, methanol, and ethanol whereas the IET for cyclohexane is its radial-symmetric counterpart. \cite{Hansen06} Since $S$ is fairly insensitive to the protein-solvent interaction potential and influenced primarily by the geometric characteristics of the polyatomic structure,\cite{Imai06} the protein can be modeled as a set of fused, neutral hard spheres. The diameter of each protein atom is set at the Lennard-Jones potential parameter $\sigma$ assigned to it.
   In the MA, $S$ is expressed as the following morphometric form:\cite{Yoshidome12,Kinoshita13,Oshima15},
\begin{eqnarray}
  \label{eq3}
  \frac{S}{k_B} = C_1 V_{ex}+C_2 A+ C_3 X+C_4 Y
\end{eqnarray}   
The EV ($V_{ex}$), solvent-accessible surface area ($A$), and integrated mean and Gaussian curvatures of the accessible surface ($X$ and $Y$, respectively) act as the geometric measures of the polyatomic structure. The four coefficients $(C_1-C_4)$, which are considered to be dependent only on the solvent species and its thermodynamic state, can then be determined beforehand by treating isolated hard-sphere solutes with various diameters. The calculation procedure is summarized below.
\begin{enumerate}
\item Calculate the solvation entropy of an isolated hard-sphere solute (SIHSS) with diameter $d_U$ using the IET. Sufficiently many different values of $d_U$, which are in the range $0.6 \le d_U/d_S \le 10$, are considered. For cyclohexane whose molecular diameter is considerably large, the range considered is $0.6 \le d_U/d_S \le 20$. For water, methanol, and ethanol, both of the translational and rotational contributions to the solvation entropy are taken into account though the former is much larger than the latter.
\item Determine $C_1-C_4$ by applying the least-squares method to the following morphometric form for isolated hard-sphere solutes:
  \begin{eqnarray}
    \label{eq4}
    \frac{{S}_{IHSS}}{k_B} = C_1 \left(\frac{4}{3}\pi R^3\right)+C_2 \left(4 \pi R^2\right) +C_3 \left(4\pi R\right) +C_4 \left(4 \pi\right) &\qquad, \qquad& R=\frac{d_{U}+d_{S}}{2}  
  \end{eqnarray}

\begin{table}[htbp]
  \caption{Four coefficients in the morphometric forms of Eqs. (3) and (4).}
  \label{tab:tab4}
  \begin{tabular}{lcccc}
    \hline \hline
    \text{Solvent}      & $C_1$ ($\mathrm{\AA}^{-3}$nm) &$C_2$ ($\mathrm{\AA}^{-2}$nm) & $C_3$ ($\mathrm{\AA}^{-1}$nm) & $C_4$ (-) \\
    \hline
    \text{Water}        &       -0.19676              &        0.04517                      &  0.25671          &    -0.35690 \\
    \text{Methanol}     &       -0.08726              &        0.04905                      & -0.03971          &     0.00918 \\
    \text{Ethanol}      &       -0.07010              &        0.03247                      & -0.02605          &     0.00075 \\
    \text{Cyclohexane}  &       -0.11111              &        0.20763                      & -0.38348          &     0.19614 \\
    \hline \hline
  \end{tabular}
\end{table}
The values of $C_1-C_4$  thus determined for water, methanol, ethanol, and cyclohexane are collected in Table \ref{tab:tab4}.
\item Calculate $V_{ex}$, $A$, $X$, and $Y$ of a protein with a prescribed structure using an extended version of Connolly’s algorithm. \cite{Connolly83,Connolly85} The $(x, y, z)$ coordinates of the center of each protein atom and its diameter $\sigma$ are served as the input data. The value of $\sigma$ is taken from the CHARMM22 force field. \cite{MacKerell98}
\item Calculate $S$ from Eq. (3) in which  $C_1-C_4$  determined in step (2) are used.
\end{enumerate}

In earlier works \cite{Oshima15,Roth06} we corroborated that the MA gives sufficiently accurate results. We tested two types of simple solvents and calculated $S$ of protein G with a variety of structures via two different routes: the three-dimensional integral equation theory (3D-IET) \cite{Ikeguchi95,Kinoshita02} and the MA combined with the radial-symmetric version of the IET. The protein polyatomic structure is explicitly treated at the atomic level by the 3D-IET. The error of the combined approach was smaller than $\pm 2\%$. Moreover, the calculation of $S$ was remarkably accelerated by the application of the MA. \cite{Roth06}
In the morphometric form hinging on the Hadwiger theorem, \cite{Likos95}  the four coefficients are equal to thermodynamic quantities of pure bulk solvent. In the case of the solvation free energy $\mu$, for instance, the first and second coefficients are the pressure $P$ and the surface tension $\gamma$, respectively. However, the Hadwiger theorem is valid only for an infinitely large solute. As argued in our earlier works, \cite{Yoshidome12,Kinoshita13,Oshima15,Kinoshita17} the form becomes problematic when it is applied to a nonpolar solute immersed in water. At $P$=1 atm, the EV term is negligibly small, and the form is approximated by $\mu \approx \gamma A$ ($\mu >0$). However, $\gamma$ becomes larger as $T$ is lowered with the result that $\mu$ increases and the hydrophobicity is strengthened. This conflicts with the experimental evidence that at low temperatures the hydrophobicity is weakened and a protein is denatured. \cite{Yoshidome12,Kinoshita13,Oshima15} No such problem arises in our form, because the four coefficients are calculated using the IET.
\subsection{Calculation of energetic component}
In the calculation of $Λ$, we choose a fully extended structure possessing the maximum number of protein-solvent hydrogen bonds (HBs) and no protein intramolecular HBs (IHBs) as the reference structure: $\Lambda=0$ for the reference structure. When a protein folds into a compact structure, many donors and acceptors (nitrogen and oxygen atoms) of the protein are buried in the interior after the break of protein-solvent HBs, but many IHBs are formed. The break and the formation lead to an increase and a decrease in energy, respectively. $\Lambda$ is calculated on the basis of this concept. We note that protein-solvent HBs are absent in cyclohexane and vacuum.
In order to determine whether an IHB is formed or not, we employ the criteria proposed by McDonald and Thornton, \cite{McDonald94} that require all the following conditions to be satisfied: The distance between centers of D and A (D is a donor and A is an acceptor) is shorter than 3.9 $\mathrm{\AA}$; the distance between centers of the hydrogen atom (H) and A is shorter than 2.5 $\mathrm{\AA}$; and the angle formed by “D−H$\cdots$A” (the dotted line signifies an HB) is larger than 90$^{\circ}$. We examine the donors and acceptors not only for backbone-backbone but also backbone-side chain and side chain-side chain IHBs. When a water-accessible surface area of a donor or an acceptor, which is calculated using Connolly’s algorithm, is smaller than 0.001 $\mathrm{\AA}$, we consider that it is buried.
\begin{figure}[htpb]
  \centering
    \includegraphics[width=.8\linewidth]{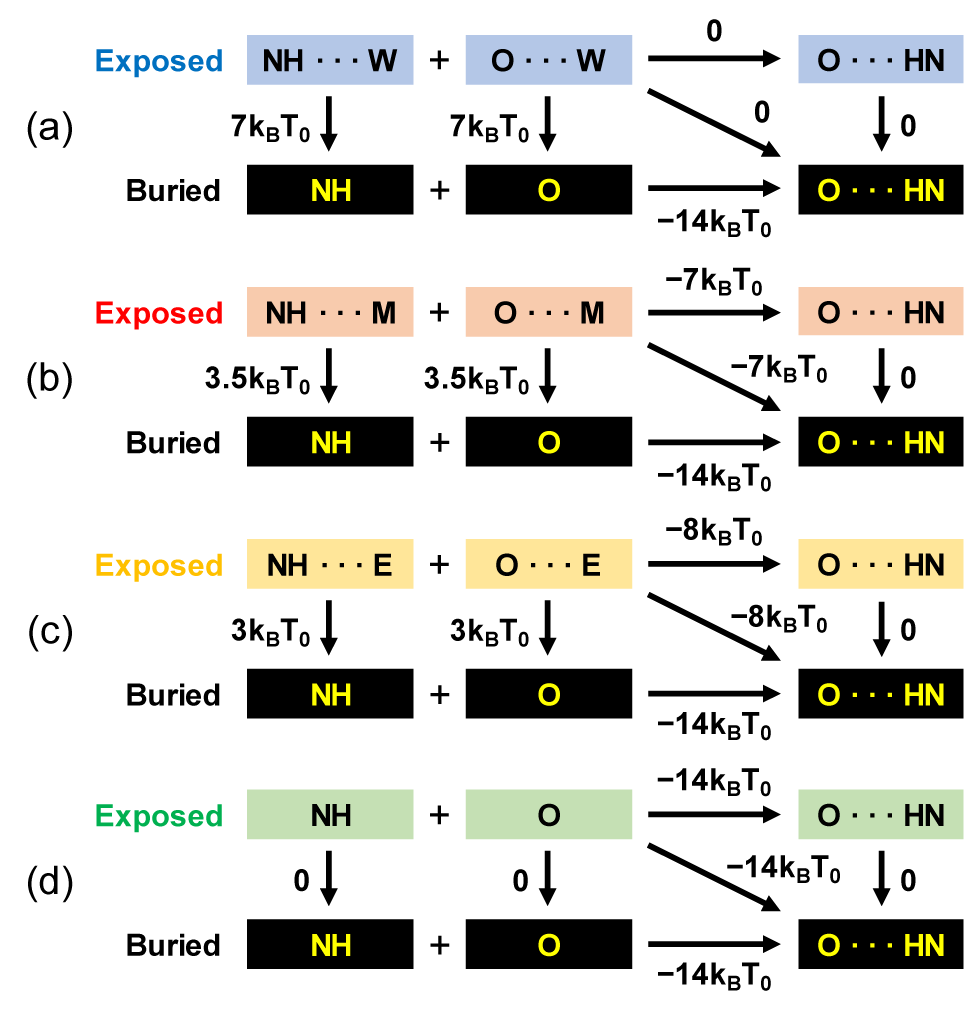}
  \caption{ Thermodynamic cycles in water (a), methanol (b), ethanol (c), and cyclohexane or vacuum (d). “$\cdots$” represents a hydrogen bond. There are four possible donor-acceptor pairs, (N, O), (O, N), (O, O), and (N, N). Here, we show the case where N is the donor and O is the acceptor. NH $\cdots$ W (Exposed) + O$\cdots$W (Exposed) $\longrightarrow$ O$\cdots$HN (Exposed) or O$\cdots$HN (Buried) in (a) is accompanied by W$\cdots$W. NH$\cdots$X (Exposed) + O$\cdots$X (Exposed) $\longrightarrow$ O$\cdots$HN (Exposed) or O$\cdots$ HN (Buried) is accompanied by X $\cdots$X (X=M (b) or E (c)). The energetic penalty $E^{*}$ is $7k_BT_0$ (a), $3.5k_BT_0$ (b), $3k_BT_0$ (c), or 0 (d) where $k_B$ is the Boltzmann constant and $T_0$=298 K.
  \label{fig:fig2}}
\end{figure}
\subsection{Hydrogen-bonding parameters for water}
 The thermodynamic cycle referring to water is illustrated in Fig. \ref{fig:fig2}(a). Upon the burial of a donor and an acceptor in the protein interior, when an IHB is formed (e.g., NH $\cdots$ W (Exposed) + O$\cdots$W (Exposed) $\longrightarrow$ O$\cdots$HN (Buried); “W” denotes an oxygen atom in a water molecule), we assume that there is no energy change occurring. When the burial of a donor or an acceptor is not accompanied by the IHB formation (e.g., NH$\cdots$W (Exposed) $\longrightarrow$ NH (Buried) and O$\cdots$W (Exposed) $\longrightarrow$ O (Buried)), an energy increase of $E^{*}$ is assumed. The formation of an IHB undergoes no energy change when the donor, acceptor, and IHB are all exposed (e.g., NH$\cdots$W (Exposed) + O $\cdots$W (Exposed)  O$\cdots$HN (Exposed)). On the other hand, the formation of an IHB within the protein interior leads to a decrease in energy of −2E (e.g., NH (Buried) + O (Buried) $\longrightarrow$ O$\cdots$HN (Buried)). It follows from the thermodynamic cycle that $−2E=−2E^{*}$. See Appendix A for more information.
$E^{*}$ is set at $7k_BT_0$ ($T_0$=298 K) (see Fig. \ref{fig:fig2}(a)) for the following reason. According to the result of quantum-chemistry calculations, if an H-acceptor pair (H is the hydrogen atom covalently bound to the donor) is completely isolated in vacuum, $−2E$ can be approximated by -$10k_BT_0$. \cite{Mitchell91}  However, the pair is in the protein interior where atoms with positive and negative partial charges are present (factor 1). Further, the decrease in energy upon protein folding occurs due to the electrostatic interaction other than hydrogen bonding as well (factor 2). Factors 1 and 2 make E smaller and larger, respectively, but factor 2 dominates for a water-soluble protein. This leads to our estimate, $−2E=14k_BT_0$ ($E^{*}=7k_BT_0$).  \cite{Hayashi17} We note that this estimate, in which all the electrostatic interactions are effectively included in the hydrogen bonding energies, has also been used successfully in our previous studies on protein folding. \cite{Yoshidome12,Hayashi17,Yoshidome09,Yasuda11,Yasuda12}
\subsection{Hydrogen-bonding parameters for methanol and ethanol}
How do the protein-water and protein intramolecular hydrogen-bonding parameters change when water is replaced by either methanol or ethanol? We first consider the energy lowering arising from the formation of protein-solvent HBs. We calculate the number of hydrogen atoms of water, methanol, or ethanol in contact with a nitrogen atom $N_N$ or with an oxygen atom $N_O$ in a peptide. \cite{Kinoshita00}  That is, we count the number of peptide nitrogen-water oxygen or peptide oxygen-water oxygen HBs. The calculations were made using the dielectrically-consistent reference interaction site model (DRISM) theory \cite{Perkyns92} combined with an all-atom model for the peptide and a modified SPC/E model \cite{Pettit82,Berendsen87} for water. The results are as follows (the numbers are normalized by those in the case where the solvent is water): $N_N$=0.454 and $N_O$=0.609 for methanol and $N_N$=0.367 and $N_O$=0.524 for ethanol. $N_N$ and $N_O$ become smaller in the order, water>methanol>ethanol. There are two reasons for this. Firstly, the number density of hydrogen atoms in the bulk liquid decreases in this order. Secondly, the steric hindrance by the hydrocarbon group in an alcohol molecule makes it more difficult for an oxygen atom in an alcohol molecule to form hydrogen bonding with a nitrogen atom or with an oxygen atom in a peptide, and since the hydrocarbon group in an ethanol molecule is bulkier than that in the methanol molecule, the steric hindrance effect for ethanol is larger.
The average value defined by $N_{av}=(N_N+N_O)/2$ is found to be 0.532 for methanol and 0.446 for ethanol. This leads to an estimate of $E^{*}=7k_BT_0 N_{av}~3.5k_BT_0$ for methanol and ~$3k_BT_0$ for ethanol (e.g., NH$\cdots$X (Exposed) $\longrightarrow$ NH (Buried) and O$\cdots$X (Exposed) $\longrightarrow$ O (Buried); X=M or E; “M” or “E” denotes the oxygen atom in a methanol or ethanol molecule). $−2E$, the energy decrease upon the formation of an IHB within the protein interior, should be independent of the solvent species. Furthermore, we assume that the burial of an IHB (e.g., O$\cdots$HN (Exposed) $\longrightarrow$ O$\cdots$HN (Buried)) accompanies no energy change as in the case of water. The protein-solvent and protein intramolecular hydrogen-bonding parameters for methanol and ethanol can be constructed as illustrated in Figs. \ref{fig:fig2} (b) and (c), respectively.
\subsection{Hydrogen-bonding parameters for cyclohexane}
Cyclohexane is a paradigmatic example of a nonpolar solvent. Even in the absence of a solvent, a donor, acceptor, and IHB are not isolated in vacuum, and significantly many protein atoms with positive and negative partial charges are close to the pair. For simplicity, we assume that the formation of an IHB always leads to an energy decrease of $−2E=−14k_BT_0$, \cite{Hayashi17} regardless of whether the donor, acceptor, or IHB is buried or not. Hence, $\Lambda$  is calculated simply by counting the number of IHBs in a protein with a prescribed structure. The resulting thermodynamic cycle is illustrated in Fig. \ref{fig:fig2}(d), which is also applied to a protein in vacuum.
   As further elaborated in Section \ref{subsec:structures} , the qualitative aspects of our conclusions are robust and not affected by the uncertainty of the hydrogen-bonding parameters set for methanol, ethanol, and cyclohexane (also see Appendix B).
\subsection{Preparation of the native and decoy structures}
\begin{figure}[htpb]
  \centering
    \includegraphics[width=.8\linewidth]{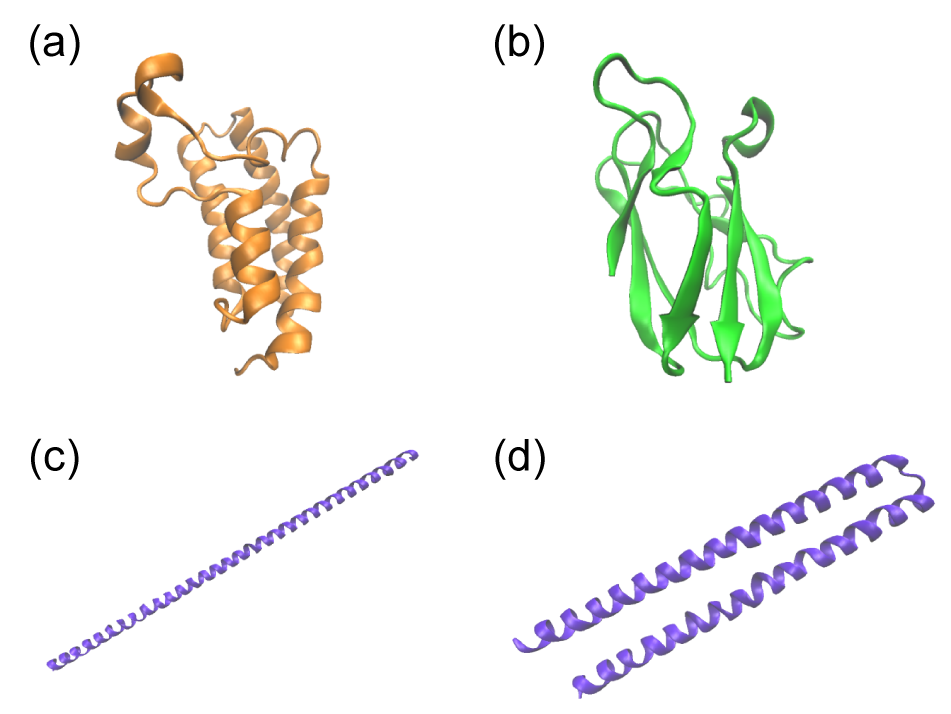}
  \caption{ (a) Native structure of CPB-bromodomain. (b) Native structure of apoplastocyanin. (c) A complete $\alpha$-helix (a single helix), “all-$\alpha$”. (d) Associated two $\alpha$-helices, “all-$\alpha$-2”.
  \label{fig:fig3}}
\end{figure}
For the NSs of CPB-bromodomain (CBP-BD) and apoplastocyanin (apoPC), we adopt the most recent data obtained by X-ray crystallography experiments (PDB CODE: 4OUF for CBP-BD45 and 2PCY for apoPC46). The NSs are shown in Figs. \ref{fig:fig3} (a) and (b), respectively. CBP-BD possesses 114 residues and apoPC 99 residues. The NS of CBP-BD is characterized by an $\alpha$-helix content of 66$\%$. The $\beta$-sheet and $\alpha$-helix contents in the NS of apoPC are 35$\%$ and 4$\%$, respectively. (In this study, the $\alpha$-helix and $\beta$-sheet contents are calculated using the DSSP program. \cite{Kabsch83})
We prepare compact decoy structures using different tools. Firstly, we exploit the 3Drobot tool, \cite{Deng16} which is an extension of the fragment assembly simulation protocol which starts from multiple structure scaffolds identified from the input structure. Unlike the other existing protocols, 3Drobot does not scarify the IHBs and the compactness of the input structure. In this study, the NS is served as the input structure and the Root-Mean-Square-Distance (RMSD) cutoff for the output decoys is set at 15 $\mathrm{\AA}$: Their values of RMSD from the input structure are smaller than 15 Å. We thus obtain 1000 decoys for each protein that are denoted as “3Drobot” hereafter. Since “3Drobot” includes decoys which are quite similar to the NS, it is suited to a stringent test of our FEF for a protein immersed in water. However, its main drawback hinges on its tendency to generate only the structures whose $\alpha$-helix and $\beta$-sheet contents are not far from those of the input structure (i.e., the NS).
\begin{figure}[htpb]
  \centering
    \includegraphics[width=.8\linewidth]{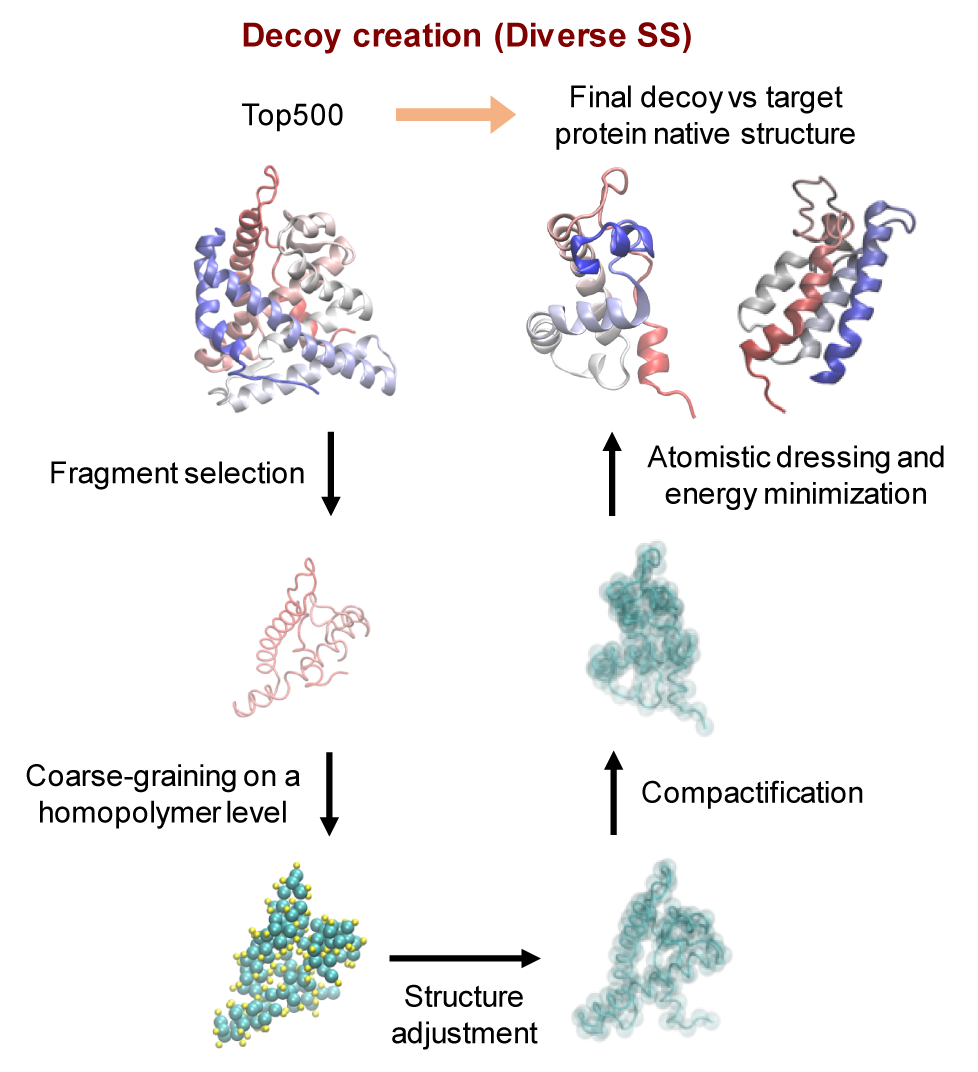}
  \caption{Schematic representation of the procedure followed to obtain the decoys in “Diverse SS”.
  \label{fig:fig4}}
\end{figure}
In order to ensure the presence of compact decoys with a wider variety of α-helix and β-sheet contents, we devise a new procedure detailed below and schematically represented in Fig. \ref{fig:fig4}. We employ the so-called “Top500 protein database” \cite{Chen10} that is regarded as a good representative of the entire PDB universe and where the template structures of a protein possess a resolution of 1.8 $\mathrm{\AA}$  or higher, with no missing atoms, a low clash score, and without unusual amino-acid substitutions. Starting from a protein structure taken from this database, we parse different fragments matching the length of the target protein. Retaining only the backbone C$_{\alpha}$ atom representation of a matching fragment, we create its homopolymer-coarse-grained representation50,51 in which every amino acid is modeled as two spherical beads, one representing the backbone atoms with a vdW radius matching that of glycine, and the other representing the side-chain atoms with a radius of 2.5 $\mathrm{\AA}$, being the approximate averaged value of the vdW radii for the 20 amino acids. Each side-chain sphere is tangent to the corresponding backbone sphere and is placed in the negative normal direction with respect to the local Frenet frame of the corresponding amino acid as detailed in a recent article. \cite{Skrbic16b} Using this coarse-grained homopolymer representation, we perform fragment compactification using a Monte Carlo Metropolis scheme aimed at minimizing the gyration radius of the chain.\cite{Skrbic16a,Skrbic16b} We employ only pivot moves with maximum angle of two degrees, in order to minimally affect the original secondary structure of the starting protein fragment. We then go back to the original all-atom protein representation employing primary sequence of the target protein and using the PULCHRA tool. \cite{Rotkiewicz08} Finally, we perform an energy minimization through a steepest descent algorithm using the GROMACS package \cite{vanderSpoel05} and discard any structure suffering unrepairable atomic clashes detected. The structures generated by this procedure are denoted as “Diverse SS” (“SS” is an abbreviation of “secondary structures”) in the rest of the paper.
   We also prepare a complete $\alpha$-helix (a single helix), “all-$\alpha$”, and associated two $\alpha$-helices, “all-$\alpha$-2”, as shown in Figs. \ref{fig:fig3} (c) and (d), respectively. The structure all-$\alpha$-2 is optimized using a molecular dynamics (MD) simulation with water modeled as a dielectric continuum.
Further, very compact structures are generated using an MD simulation in vacuum. The initial structure is either the NS or all-$\alpha$, and only the protein intramolecular vdW interaction and the bonded energy are taken into consideration. The structures generated via this method are identified as “MD (vdW)”, with some of them possessing little secondary structures. Last, we generate a set of structures classified as “MD (GB)”. Many of them are characterized by quite high $\alpha$-helix contents. They are generated using an MD simulation with water modeled as a dielectric continuum. For CBP-BD, the initial structure is either all-$\alpha$-2 or one of the two structures chosen from “3Drobot” as those which are quite stable in vacuum (i.e., in terms of IHBs). For apoPC, the initial structure is either all-$\alpha$-2 or one of the two structures chosen from “Diverse SS” as those with relatively higher $\alpha$-helix contents.
   The slight, unrealistic overlaps of the constituent atoms occurring in a protein structure are removed by the local minimization of the energy function using the CHARMM54 and MMTSB55 packages based on the CHARMM22 force field \cite{MacKerell98} combined with the CMAP correction \cite{Mackerell04} and the GBMV implicit solvent model. \cite{Lee03,Chocholousova06}
We believe that the decoy structures generated in this study are sufficiently many and diverse for the following reasons: (1) 3Drobot used in this study was developed especially for testing a free-energy function in terms of its performance of discriminating the NS (i.e., the structure stabilized in aqueous solution under physiological conditions) from a number of misfolded decoys. The structures which resemble the NS very much are included in the decoys generated by 3Drobot; (2) in addition to 3Drobot, using our own new procedure, we generate decoys with a wide variety of α-helix and β-sheet contents and values of the RMSD from the NS; and (3) further, using MD simulations, we generate significantly many structures. One of the MD simulations is performed in vacuum. Only the protein intramolecular vdW interaction and the bonded energy are taken into consideration (i.e., all of the partial charges of protein atoms are shut off). This type of simulation tends to generate more compact structures with fewer intramolecular hydrogen bonds and less content of the secondary structures. The other MD simulation is undertaken so that structures with quite high $\alpha$-helix contents can be generated. The decoys in (2) and (3) are necessitated because we consider not only water but also methanol, ethanol, and cyclohexane as the solvent.
\section{Results for an $\alpha$-helix-rich protein: CPB-BD}
We now proceed to assess the structural stability of a protein in water, methanol, ethanol, cyclohexane, and vacuum by contrasting the FEF of its NS with that of all the misfolded decoys generated by the procedure explained above. We start with CPB-BD, an $\alpha$-helix-rich protein. We discriminate between the entropic and energetic components of the FEF to underpin the relative role of each component. $\Delta F<0$ ,$-\Delta S>0$, or $\Delta \Lambda<0$ identifies a particular decoy that is more stable than the NS in each solvent with respect to the FEF, entropic component, or energetic one.
\begin{enumerate}
\item [A.] In water \\
\begin{figure}[htpb]
  \centering
    \includegraphics[width=.8\linewidth]{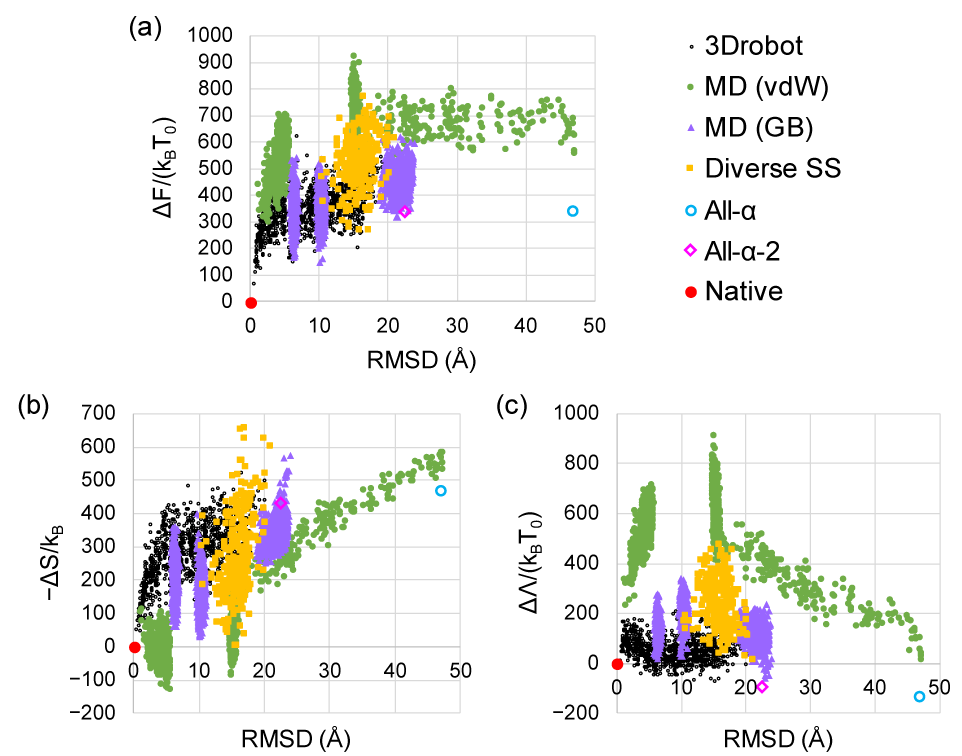}
  \caption{Relation between the RMSD from the native structure in terms of the C$_{\alpha}$ atoms and $\Delta F/(k_BT_0)$ (a), $−\Delta S/k_B$ (b), or $\Delta \Lambda/(k_BT_0)$ (c): CPB-bromodomain (CPB-BD) in water is considered.
  \label{fig:fig5}}
\end{figure}
\begin{figure}[htpb]
  \centering
    \includegraphics[width=.8\linewidth]{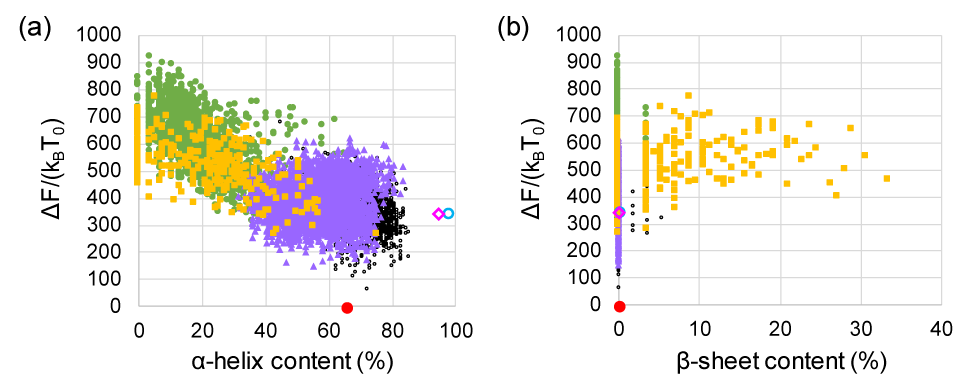}
  \caption{$\Delta F/(k_BT_0)$ plotted against $\alpha$-helix content (a) or $\beta$-sheet content (b): CPB-BD in water is considered. See Fig. \ref{fig:fig5} for the class of structures (“3Drobot”, “MD (vdW)”, etc.) represented by each key.
  \label{fig:fig6}}
\end{figure}
   Figure \ref{fig:fig5} shows the relation between the RMSD from the NS in terms of the C$_{\alpha}$ atoms and $\Delta F/(k_BT_0)$ (a), $−\Delta S/k_B$ (b), or $\Delta \Lambda/(k_BT_0)$ (c). Interestingly, there are significantly many structures which are more stable than the NS with respect to one of $\Delta \Lambda/(k_BT_0)$ and $−\Delta S/k_B$. However, the NS is always the optimal choice in terms of the sum of the two (i.e. in terms of $\Delta F/(k_BT_0)$) and is thus the most stable. The two components are often opposing, which is exemplified in the plots for “MD (vdW)”. “MD (vdW)” includes the structures which are extremely compact and entropically more stable than the NS but suffer the lack of IHBs causing the energetic instability. Though there is a tendency that ΔF decreases and the stability becomes higher as the $\alpha$-helix content increases as observed in Fig. \ref{fig:fig6}(a), all-$\alpha$ and all-$\alpha$-2 are much less stable than the NS. As expected, all-$\alpha$ is the most stable with respect to $\Delta \Lambda$. It is interesting that for the $\alpha$-helix-rich protein $\Delta F$ is less correlated with the $\beta$-sheet content (see Fig. \ref{fig:fig6}(b)).
 \item [B.] In methanol or ethanol \\
\begin{figure}[htpb]
  \centering
    \includegraphics[width=.8\linewidth]{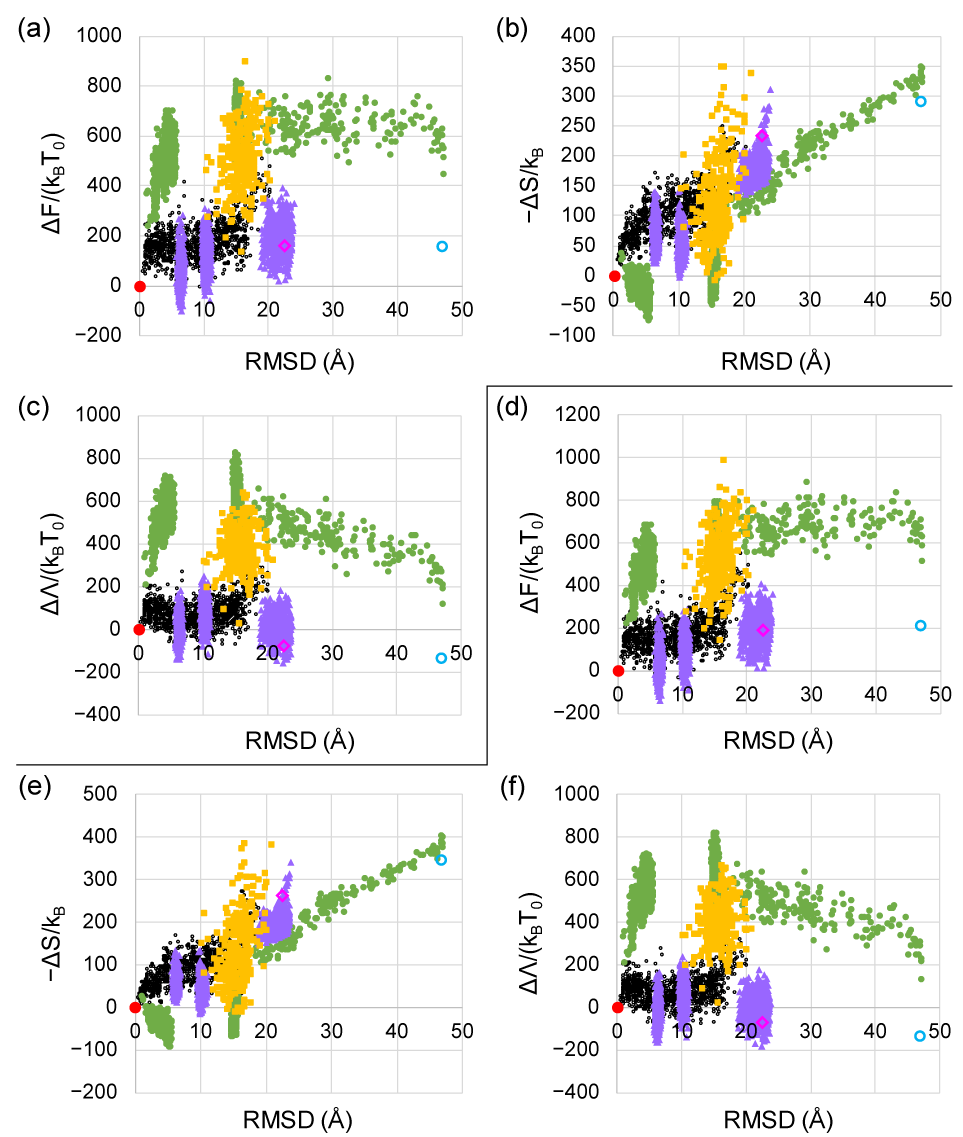}
  \caption{Relation between the RMSD from the native structure in terms of the  C$_{\alpha}$ atoms and $\Delta F/(k_B T_0)$  (a), $−\Delta S/k_B$ (b), or $\Delta \Lambda/(k_BT_0)$ (c): CPB-BD in methanol is considered. Relation between the RMSD and $\Delta F/(k_BT_0)$ (d), $−\Delta S/k_B$ (e), or $\Delta \Lambda /(k_BT_0)$ (f): CPB-BD in ethanol is considered. See Fig. 5 for the class of structures (“3Drobot”, “MD (vdW)”, etc.) represented by each key.
  \label{fig:fig7}}
\end{figure}
\begin{figure}[htpb]
  \centering
    \includegraphics[width=.8\linewidth]{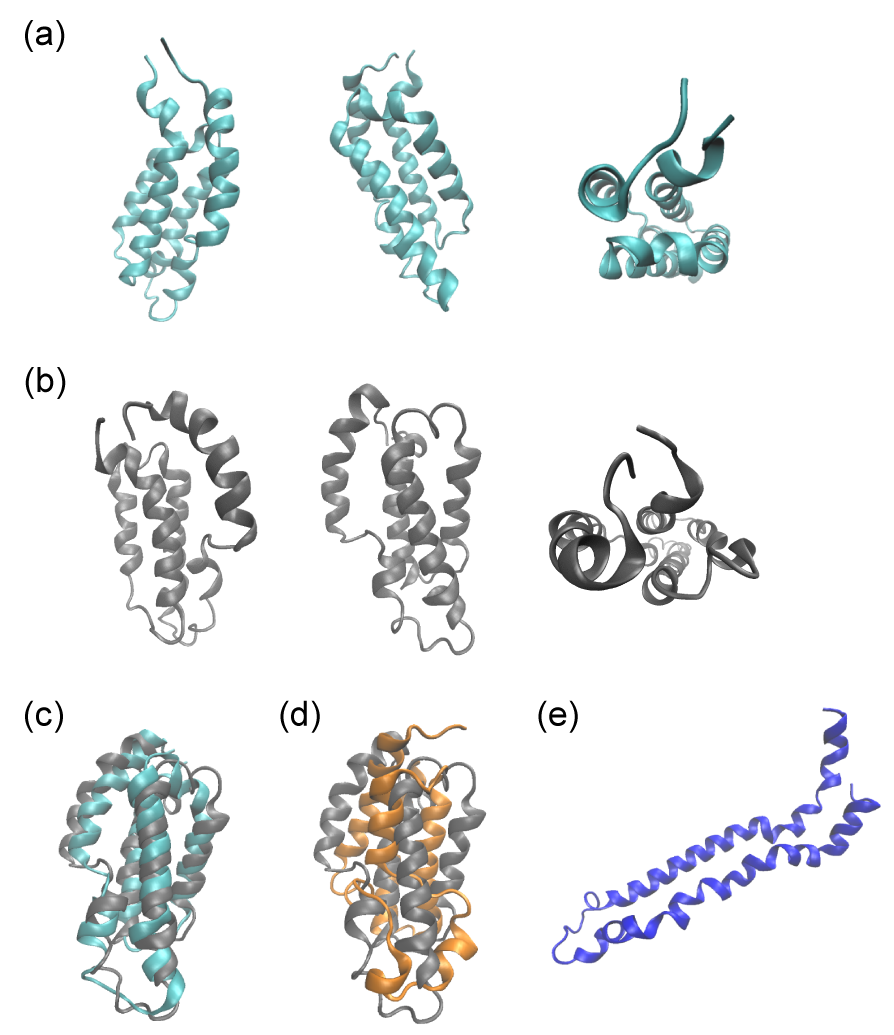}
  \caption{(a) The most stable structure of CPB-BD in methanol and ethanol (colored in light blue). It is viewed from three different angles. (b) The most stable structure of CPB-BD in cyclohexane. It is viewed from three different angles. (c) Superimposition of the most stable structures of CPB-BD in methanol and ethanol (colored in light blue) and in cyclohexane (colored in grey). (d) Superimposition of the most stable structure of CPB-BD in cyclohexane and the native structure (colored in orange). (e) The most stable structure of CPB-BD in vacuum.
  \label{fig:fig8}}
\end{figure}
\begin{figure}[htpb]
  \centering
    \includegraphics[width=.8\linewidth]{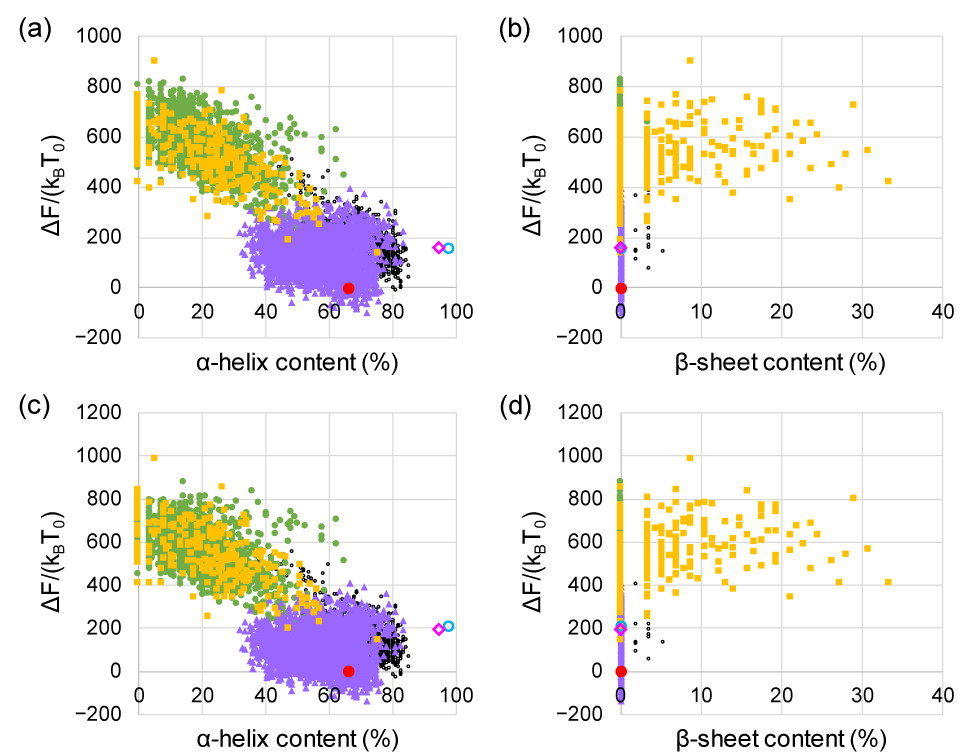}
  \caption{$\Delta F/(k_BT_0)$ plotted against $\alpha$-helix content in methanol (a) or ethanol (c). $\Delta F/(k_BT_0)$ plotted against $\beta$-sheet content in methanol (b) or ethanol (d). The protein considered is CPB-BD. See Fig. 5 for the class of structures (“3Drobot”, “MD (vdW)”, etc.) represented by each key.
  \label{fig:fig9}}
\end{figure}
   Figure \ref{fig:fig7} reports $\Delta F/(k_BT_0)$ (a), $−\Delta S/k_B$ (b), or $\Delta \Lambda /(k_BT_0)$ (c) as a function of the RMSD and for methanol. The analog plots for ethanol are shown in (d), (e), and (f), respectively. Overall, the results for methanol and ethanol look quite similar. Significantly many structures are more stable than the NS in terms of $\Delta F /(k_BT_0)$, for which the energetic component is mainly responsible. The same structure depicted in Fig. \ref{fig:fig8}(a) is found to be the most stable in the two alcohols. It is characterized by an association of four $\alpha$-helices with an $\alpha$-helix content of 72 $\%$. It is significantly different from the NS (see Fig. \ref{fig:fig3}). We look at the second most stable one, but this structure and the structure in Fig. \ref{fig:fig8}(a) share the same characteristics expressed as “associated $\alpha$-helices”. It is possible that there is a structure which is even more stable than the two structures but missing in the decoy structures generated. However, we believe that it also comprises associated $\alpha$-helices. As observed in Figs. \ref{fig:fig9} (a) and (c), the stability tends to become higher as the $\alpha$-helix content increases for both methanol and ethanol. However, neither all-$\alpha$-2 nor the all-$\alpha$ is very stable. $\Delta F$ is less correlated with the $\beta$-sheet content (see Figs. \ref{fig:fig9}(b) and (d)).
 \item [C.] In cyclohexane \\
\begin{figure}[htpb]
  \centering
    \includegraphics[width=.8\linewidth]{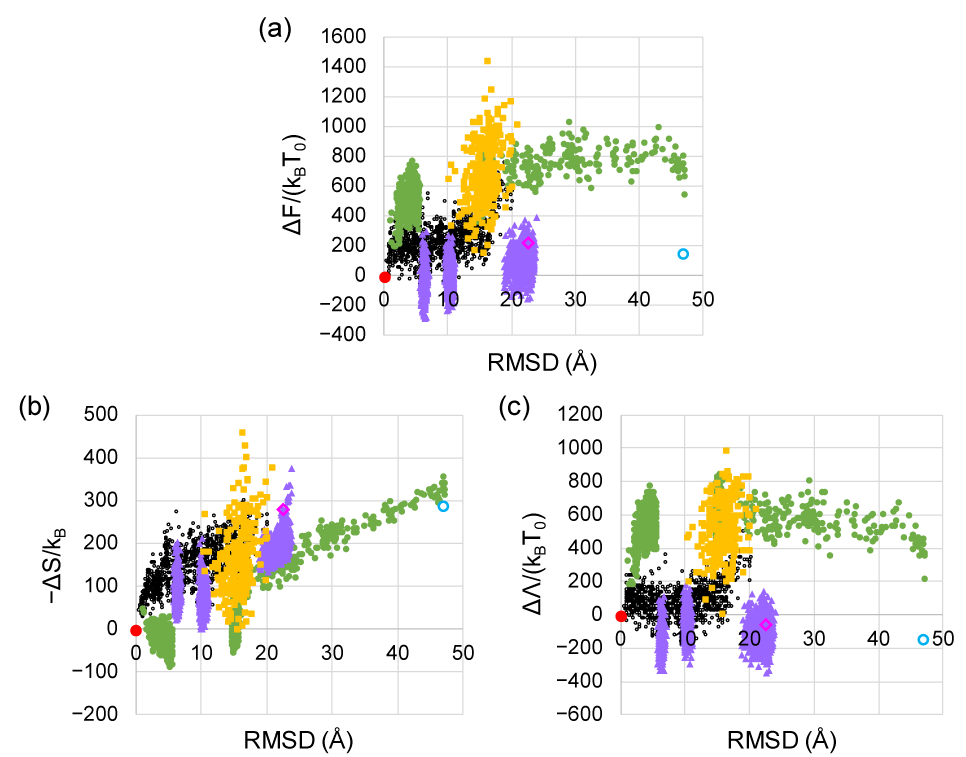}
  \caption{Relation between the RMSD from the native structure in terms of the C$_{\alpha}$ atoms and $\Delta F/(k_BT_0)$ (a), $−\Delta S/k_B$ (b), or $\Delta \Lambda/(k_BT_0)$ (c): CPB-BD in cyclohexane is considered. See Fig. \ref{fig:fig5} for the class of structures (“3Drobot”, “MD (vdW)”, etc.) represented by each key.
  \label{fig:fig10}}
\end{figure}
   In Fig. \ref{fig:fig10}, we plot $\Delta F/(k_BT_0)$ (a), $−\Delta S/k_B$ (b), or $\Delta \Lambda /(k_BT_0)$ (c) against the RMSD. The results observed in Fig. \ref{fig:fig10} are qualitatively similar to those obtained for the two alcohols. The structure identified as the most stable one is shown in Fig. \ref{fig:fig8}(b). It is characterized by an association of four $\alpha$-helices with an $\alpha$-helix content of 61$\%$. As compared in Fig. \ref{fig:fig8}(c), the most stable structures in methanol, ethanol, and cyclohexane are similar to one another but somewhat different from the NS (see Fig. \ref{fig:fig8} (d)). There is a tendency that the stability becomes higher as the $\alpha$-helix content increases, but it is less correlated with the $\beta$-sheet content (the plots are not shown).
\item [D.] In vacuum \\
There is only the energetic component for vacuum. The relation between the RMSD and $\Delta F/(k_BT_0)$ is obtained by replacing $\Delta \Lambda /(k_BT_0)$ in Fig. \ref{fig:fig10}(c) by $\Delta F /(k_BT_0)$. The most stable structure is shown in Fig. \ref{fig:fig8}(e). It is characterized by an association of two $\alpha$-helices with an $\alpha$-helix content of 71$\%$. Again, there is a tendency that the stability becomes higher as the $\alpha$-helix content increases, but it is less correlated with the $\beta$-sheet content (the plots are not shown).
\end{enumerate}
\section{Results for a $\beta$-sheet-rich protein: apoPC}
We then consider the results for the mirrored calculation for apoPC, a $\beta$-sheet-rich protein.
\begin{enumerate}
\item [A.] In water \\
\begin{figure}[htpb]
  \centering
    \includegraphics[width=.8\linewidth]{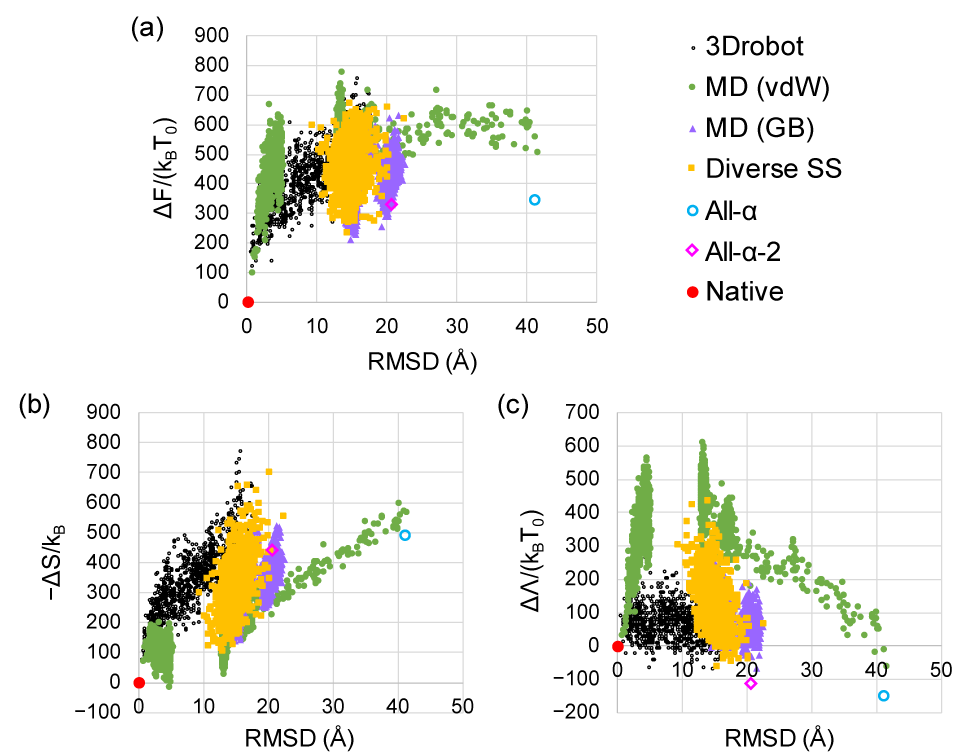}
  \caption{Relation between the RMSD from the native structure in terms of the C$_{\alpha}$ atoms and $\Delta F/(k_BT_0)$ (a), $−\Delta S/k_B$ (b), or $\Delta \Lambda /(k_BT_0)$ (c): apoplastocyanin (apoPC) in water is considered.
  \label{fig:fig11}}
\end{figure}
\begin{figure}[htpb]
  \centering
    \includegraphics[width=.8\linewidth]{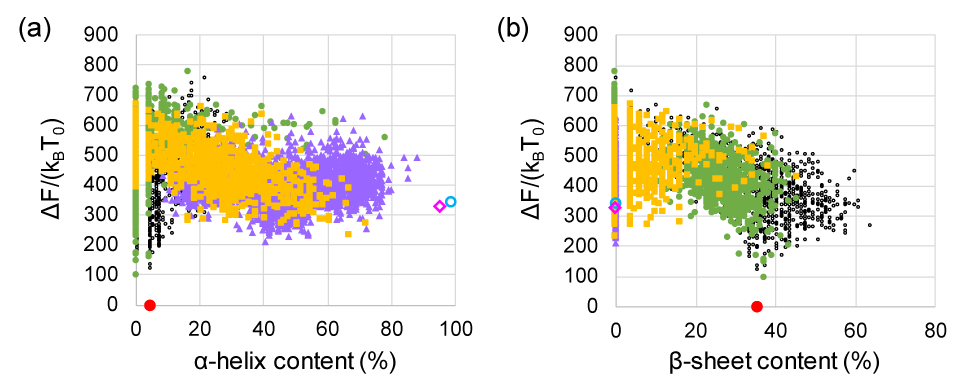}
  \caption{$\Delta F/(k_BT_0)$ plotted against $\alpha$-helix content (a) or $\beta$-sheet content (b): apoPC in water is considered. See Fig. \ref{fig:fig11} for the class of structures (“3Drobot”, “MD (vdW)”, etc.) represented by each key.
  \label{fig:fig12}}
\end{figure}
   Figure \ref{fig:fig11}  shows the relation between the RMSD from the NS in terms of the C$_{\alpha}$ atoms and $\Delta F/(k_BT_0)$ (a), $−\Delta S/k_B$ (b), or $\Delta \Lambda /(k_BT_0)$ (c). The NS is again correctly identified as the most stable structure using $\Delta F/(k_B T_0)$, though there are significantly many structures which are more stable than the NS with respect to one of the energetic and entropic components. In contrast to the $\alpha$-helix-rich protein, the structure of the $\beta$-sheet-rich protein tends to become more stable as the β-sheet content increases whereas the stability is less correlated with the $\alpha$-helix content (see Figs. \ref{fig:fig12}(a) and (b)).
 \item [B.] In methanol or ethanol \\
\begin{figure}[htpb]
  \centering
    \includegraphics[width=.8\linewidth]{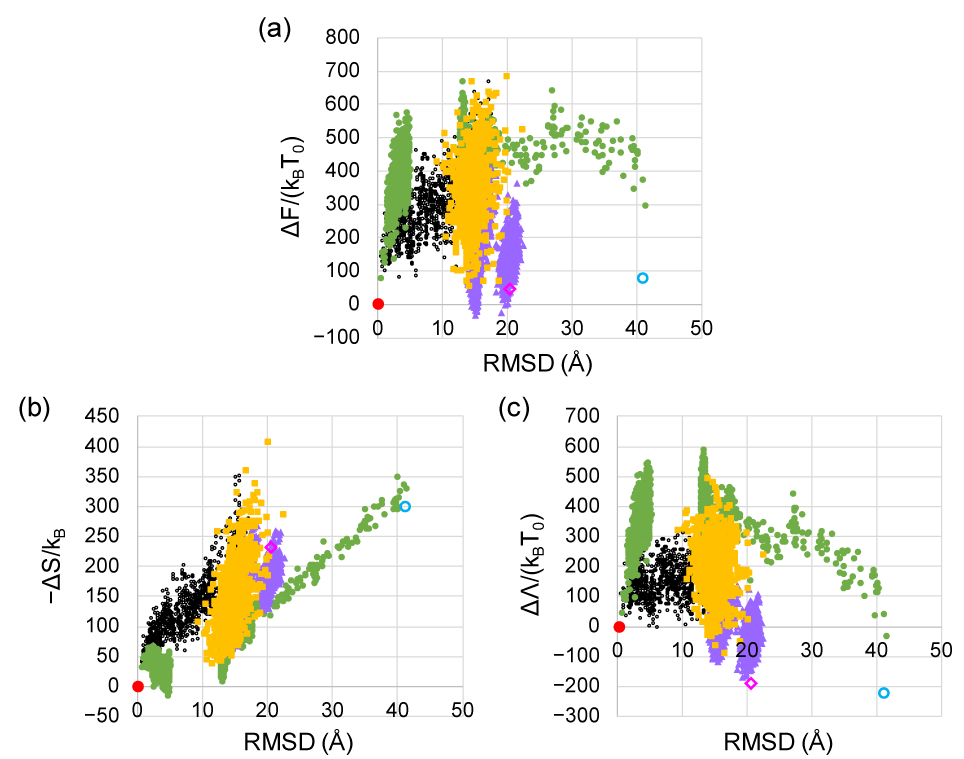}
  \caption{Relation between the RMSD from the native structure in terms of the C$_{\alpha}$ atoms and $\Delta F/(k_BT_0)$ (a), $-\Delta S/k_B$ (b), or $\Delta \Lambda/(k_BT_0)$ (c): apoPC in methanol is considered. See Fig. \ref{fig:fig11} for the class of structures (“3Drobot”, “MD (vdW)”, etc.) represented by each key.
  \label{fig:fig13}}
\end{figure}
\begin{figure}[htpb]
  \centering
    \includegraphics[width=.8\linewidth]{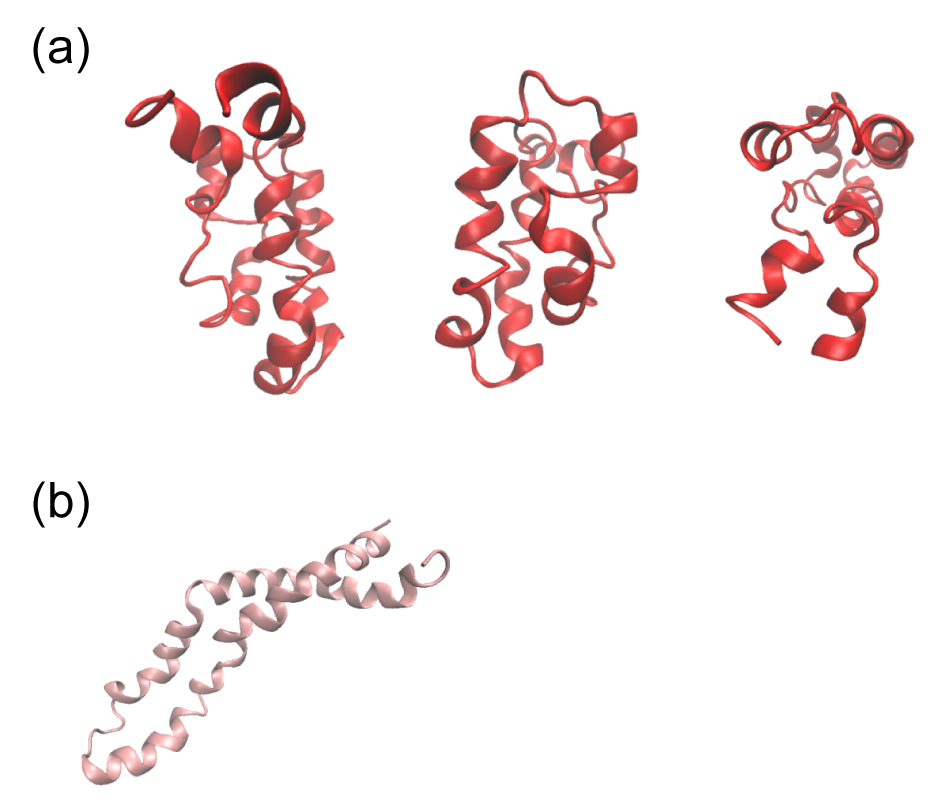}
  \caption{(a) The most stable structure of apoPC in methanol, ethanol, and cyclohexane. It is viewed from three different angles. (b) The most stable structure of apoPC in vacuum.
  \label{fig:fig14}}
\end{figure}
\begin{figure}[htpb]
  \centering
    \includegraphics[width=.8\linewidth]{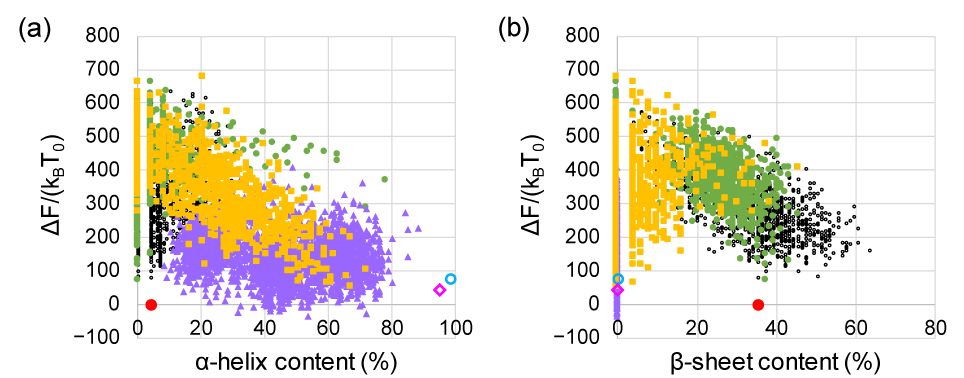}
  \caption{ $\Delta F/(k_BT_0)$ plotted against $\alpha$-helix content (a) or $\beta$-sheet content (b): apoPC in methanol is considered. See Fig. \ref{fig:fig11} for the class of structures (“3Drobot”, “MD (vdW)”, etc.) represented by each key.
  \label{fig:fig15}}
\end{figure}
   Since the results for methanol and ethanol are qualitatively the same, we present only those for methanol. As observed in Figs. \ref{fig:fig13}(a), (b), and (c) where $\Delta F /(k_BT_0)$,$ −\Delta S/k_B$, and $\Delta \Lambda /(k_BT_0)$ are plotted against the RMSD, respectively, there are structures which are more stable than the NS in terms of $\Delta F/(k_BT_0)$, which is attributed to the energetic component. The most stable structure is illustrated in Fig. \ref{fig:fig14}(a). We note that the same structure is identified as the most stable one in ethanol. It is characterized by an association of four α-helices with an α-helix content of 54$\%$. It is interesting that $\Delta F$ is almost equally correlated with the $\alpha$-helix and $\beta$-sheet contents (see Figs. \ref{fig:fig15} (a) and (b)).
\item [C.] In cyclohexane \\
   As in the case of CPB-BD, the results are qualitatively similar to those for the two alcohols. The structure identified as the most stable one in the two alcohols is also the most stable in cyclohexane (see Fig. \ref{fig:fig14}(a)). Though the NS is rich in the $\beta$-sheet, the structure stabilized in methanol, ethanol, and cyclohexane is rich in the $\alpha$-helix. $\Delta F$ is almost equally correlated with the $\alpha$-helix and $\beta$-sheet contents (the plots are not shown).
\item [D.] In vacuum \\
  The most stable structure in vacuum is shown in Fig. \ref{fig:fig14}(b). It is characterized by an association of two $\alpha$-helices with an $\alpha$-helix content of 71$\%$. $\Delta F$  is almost equally correlated with the $\alpha$-helix and $\beta$-sheet contents (the plots are not shown).
\end{enumerate}
\section{Discussion}
\subsection{Structures of a protein stabilized in different model solvents}
\label{subsec:structures}
The formation of the $\alpha$-helix or $\beta$-sheet leads not only to a solvent-entropy gain (see Fig. \ref{fig:fig1}(a) and (b)) but also to the compensation of break of protein-solvent HBs by the assurance of IHBs. For this reason, these secondary structures are fundamental units most favored in the NS. In general, the $\alpha$-helix is capable of forming more IHBs than the $\beta$-sheet, because in the latter significantly many donors and acceptors remain unpaired. Therefore, when the energetic component dominates, the $\alpha$-helix is more favorable than the $\beta$-sheet and a structure possessing the highest possible $\alpha$-helix content is stabilized.
In the absence of the solvent-entropy effect (i.e., in vacuum), the structure with the maximum number of IHBs is highly favorable. In terms of the number of backbone-backbone IHBs, all-α is the most stable. However, backbone-side chain and side chain-side chain IHBs can also be formed, which makes a significant contribution to the structural stability. We find that an association of two α-helices is more stable than all-α. It is even more stable than all-α-2 due to a more favorable arrangement of the two α-helices (compare Fig. \ref{fig:fig3}(d) with Figs. \ref{fig:fig8}(e) and \ref{fig:fig14}(b)). Remarkably, this result is independent of the amino-acid sequence and therefore universal.
   We then consider a protein immersed in cyclohexane, ethanol, and methanol. The strength of the energetic component follows the expected order, “cyclohexane>ethanol>methanol” (see Fig. \ref{fig:fig2}). The energetic penalty $E^{*}$ caused by the break of a protein-solvent HB, which vanishes in cyclohexane, is larger in methanol than in ethanol, resulting in the above ordering (see Fig. \ref{fig:fig2}). The differences among the three solvents in terms of the strength of entropic EV effect are rather small. This can be rationalized as follows. The EV effect becomes stronger as dS decreases and/or $\eta_{S}$ increases. The values of $d_S$ and $\eta_S$ of the three solvents are given in Table \ref{tab:tab3}. According to the AO theory,\cite{Asakura54,Asakura58}which accounts for only the EV term of the solute-solvent pair correlation \cite{Hansen06} but can be used in an approximate discussion, the strength is in proportion to $\eta_S/d_S$  (0.939, 0.873, and 0.914 for methanol, ethanol, and cyclohexane, respectively). Qualitatively, all the three solvents possess the common characteristic that the energetic component is stronger than the entropic one but the latter also contributes significantly to the structural stability. With respect to the solvent-entropy effect, close packing of side chains is the most important. Since this effect is moderately strong in methanol, ethanol, and cyclohexane, the close side-chain packing is also necessitated. A good solution is to construct an association of more than two $\alpha$-helices achieving overall closer side-chain packing and more reduction of the total EV: We find that an association of four $\alpha$-helices (see Figs. \ref{fig:fig8}(a), (b), and \ref{fig:fig14}(a)) is quite stable.
As discussed above, the structures stabilized in vacuum, cyclohexane, ethanol, and methanol can all be characterized by an association of $\alpha$-helices as structural units, which is suggestive that the qualitative aspects of our conclusions are not affected by the uncertainty of the hydrogen-bonding parameters set for these three solvents. However, the number of the units is smaller in vacuum than in methanol, ethanol, and cyclohexane. This result is independent of the amino-acid sequence. Even the $\beta$-sheet protein, apoPC, would change its structure to an association of a-helices if placed in these solvent environments.
   In water, the effect of the entropic component is exceptionally strong. Thanks to the hydrogen-bonding network, water exists as a dense liquid at ambient temperature and pressure despite its quite a small molecular size ($\eta_S/d_S$=1.37). More importantly, in our earlier works \cite{Oshima15,Hayashi17,Yoshidome09,Yasuda11} we demonstrated that the effect is considerably stronger in water than in the hard-sphere solvent for which $d_S$ and $\eta_S$ are set at the values of water. In other words, the dependence of the solvation entropy of a protein on its structure becomes largest when the solvent is water, which originates from the strong water-water attractive interaction potential. On the other hand, the energetic penalty $E^{*}$ is quite large, much larger than that of methanol (see Fig. \ref{fig:fig2}). It follows that in water the entropic component is at least as strong as the energetic one. It is required that the secondary structures (preferably, the $\alpha$-helix) be formed as much as possible, but close packing of side chains is also imperative. The close packing cannot always be achievable with a structure possessing a high $\alpha$-helix content depending on the amino-acid sequence. A good example is apoPC for which the $\beta$-sheet is preferentially chosen to achieve the close side-chain packing.
\subsection{Comparison with experimental observations}
   According to the experimental observations, \cite{Hirota98} alcohol induces a protein to form α-helices and the helical structure induced by alcohol is independent of the alcohol species. Our results are in good accord with these observations.
   For proteins in nonpolar solvents, two additional interesting features are known in the literature: \cite{Pace04,Griebenow96,Klibanov01}
\begin{enumerate}
\item[(1)] An enzyme is not soluble in a nonpolar solvent but forms suspensions (some water molecules are retained on the enzyme surface). The enzyme NS, which is not collapsed, is considerably more thermostable than in water.
\item[(2)] According to the experimental studies examining enzymes in aqueous-organic mixtures,\cite{Griebenow96} an enzyme is denatured. However, as stated in (1), when the enzyme is introduced into a nonpolar solvent, its NS is retained. The reason for this counter-intuitive behavior is that in the absence of water, enzymes are very rigid.
  \end{enumerate}
Thus, it is likely that the NS of a protein is retained even after it is introduced into a nonpolar solvent. This can be interpreted as follows. In the nonpolar solvent, the NS becomes highly stable (more stable than in water), because it is very difficult to break the protein IHBs already formed in the NS. That is, the NS is a metastable state: There is a very high free-energy barrier for the protein to overcome to reach the most stable structure that should be characterized by associated a-helices. It is also well known that the addition of a small amount of water to the protein-nonpolar solvent system (the environment is still far from the aqueous one) makes a protein more flexible and denatured.\cite{Griebenow96,Klibanov01} In this case, the protein can overcome the free-energy barrier for the denaturation because the energy increase due to the break of IHBs can be compensated by the energy decrease brought by the formation of protein-water HBs. The protein can change its structure to the most stable one. Thus, features (1) and (2) are consistently interpretable though they are not actually reproduced in our calculations.

\section{Conclusions}
   We have investigated the structures of two proteins, CPB-bromodomain (CBP-BD) \cite{Mujtaba04} and apoplastocyanin (apoPC), \cite{Garret84}  stabilized in model water, methanol, ethanol, cyclohexane, and vacuum using our free-energy function (FEF). \cite{Hayashi17,Yoshidome09,Yasuda11} The structure stabilized in aqueous solution under the physiological condition is referred to as “native structure (NS)”. The NS of CBP-BD possesses an $\alpha$-helix content of 66$\%$. That of apoPC possesses $\beta$-sheet and $\alpha$-helix contents of 35$\%$ and 4$\%$, respectively. A water molecule is modeled as a hard sphere in which a point dipole and a point quadrupole of tetrahedral symmetry are embedded.\cite{Kusalik88a,Kusalik88b}  The model of a methanol or ethanol molecule is a hard sphere in which only a point dipole is embedded.\cite{Kinoshita91} A cyclohexane molecule is modeled as a neutral hard sphere. The transition to a compact structure of a protein is accompanied by the break of protein-solvent hydrogen bonds (HBs), formation of protein intramolecular HBs (IHBs), and recovery of solvent-solvent HBs. These are taken into account in the energetic component of the FEF. Protein-solvent and solvent-solvent HBs are not present in cyclohexane and vacuum. The hydrogen-bonding parameters employed in calculating the energetic component have newly been determined for methanol and ethanol. The structural transition is also accompanied by a solvent-entropy gain except in vacuum. The diameter, packing fraction, and multipoles are parameterized to match the basic properties of each solvent. The solvation entropy of a protein with a prescribed structure is calculated using a radial-symmetric \cite{Hansen06} or angle-dependent \cite{Kinoshita08,Kusalik88a,Kusalik88b,Kinoshita96,Cann97} integral equation theory combined with our morphometric approach. \cite{Yoshidome12,Oshima15,Koning04,Roth06} 
   For a protein, it is important to form as many IHBs as possible (requirement 1). This requirement becomes stronger in the order: “vacuum=cyclohexane>ethanol>methanol>water”. It is also important to keep the solvent entropy as high as possible except in vacuum (requirement 2). This requirement becomes stronger in the order: “water>methanolethanolcyclohexane”. The optimally stabilized structure is determined by the competition of requirements 1 and 2. However, requirement 1 dominates in all solvents except water, and the priority is then given not to the close side-chain packing but to the formation of a maximum number of IHBs. In this case, the $\alpha$-helix is more favorable than the $\beta$-sheet because the donors and acceptors left unbounded unavoidably remain in the latter. Therefore, the most stable structure is characterized by an association of $\alpha$-helices with a high $\alpha$-helix content.
   For both of CBP-BD and apoPC, the most stable structures in methanol and ethanol are the same and can be characterized by an association of a-helices as structural units. The most stable structure in cyclohexane is also similar. In vacuum, the structure possessing the maximum number of IHBs is stabilized. It is also characterized by an association of $\alpha$-helices as structural units. Since the solvent-entropy effect certainly works in methanol, ethanol, and cyclohexane, the number of the units is made larger than in vacuum so that the close side-chain packing can be achieved to some extent. The qualitative characteristics of the structures stabilized in vacuum and the three solvents are the same for both CBP-BD and apoPC and therefore almost independent of the amino-acid sequence. They are somewhat different from the NS even for the $\alpha$-helix-rich protein CBP-BD because of the crucial role played by the close side-chain packing in water.
   In water, requirement 2 is as strong as or even stronger than requirement 1. In particular, the close packing of side chains is essential. A high $\alpha$-helix content is not necessarily suited to the achievement of sufficiently close packing. There are many cases where the $\beta$-sheet is the preferentially selected elemental unit. The contents of $\alpha$-helix, $\beta$-sheet, and total secondary structure are optimized in terms of the interplay of the energetic and entropic effects. This explains why, only in water, a variety of structures are stabilized depending on the amino-acid sequence. It has been verified that our FEF is capable of discriminating the NS from a number of decoy structures as the one for which the FEF becomes lowest.
In this study, we consider CPB-BD and apoPC. In our preceding work, \cite{Hayashi17} we considered protein G with 56 residues possessing $\alpha$-helix and $\beta$-sheet contents of 27$\%$ and 39$\%$, respectively. Importantly, we considered 133 proteins in an earlier work. \cite{Yasuda11} Thus, the number of proteins we have tested is 136. Using the same FEF, we have been successful in discriminating the NS from a number of decoys for 131 proteins. We were unsuccessful for the other 5 proteins. \cite{Yasuda11} However, this nonsuccess can be justified as follows. For two of them, the structures stabilized under acidic conditions (pH=3.5 and 4.5) are regarded as the NSs. They should be significantly different from the true NSs. For any of the other three proteins, portions of the terminus sides are removed and a significantly high percentage of the secondary structures is lost. In summary, as long as the NS is not unrealistic, we have always been successful in the discrimination.
A membrane protein is in the environment that is similar to a nonpolar solvent like cyclohexane. The fact that its stabilized structure is usually characterized by associated $\alpha$-helices is in good accord with the result mentioned above. There is another type of membrane-protein structure featuring the $\beta$-barrel. In the $\beta$-barrel there are very few donors and acceptors without forming IHBs unlike in the $\beta$-sheet. In this sense, the $\beta$-barrel is more like the $\alpha$-helix. It is intriguing to explore if the $\beta$-barrel structure becomes more stable than an association of $\beta$-helices when the amino-acid sequence of a $\beta$-barrel protein is used in the calculation. It is definite that the solvent-entropy effect plays a pivotal role in distinguishing the two types of structures through the achievement of close packing of side chains. For a $\beta$-barrel protein, an association of $\alpha$-helices should be much more unfavorable from the entropic point of view. Work in this direction is in progress.
\acknowledgments{One of the authors (M. K.) developed the computer program for the MA with R. Roth and Y. Harano. This work was supported by Grant-in-Aid for Scientific Research (B) (No. 17H03663) from Japan Society for the Promotion of Science (JSPS) to M. K. and by MIUR PRIN-COFIN2010-2011 (contract 2010LKE4CC) to A. G. The use of the SCSCF multiprocessor cluster at the Università Ca’ Foscari Venezia is gratefully acknowledged.}


\appendix
\section{Comment of consistency with all-atom computer simulation studies}
Since a rather simplified set of models are employed in this study, it may be worthwhile to comment on the consistency with all-atom computer simulation studies for the following two representative points: the importance of the translational entropy of water and the reliability of our physical picture illustrated in Fig. 2(a).
Using a Monte Carlo simulation, Paulaitis and coworkers \cite{Lazaridis92,Ashbaugh96} studied relative magnitudes of the translational and orientational restrictions contributing to the solvation entropy of a nonpolar solute inserted into water. They considered only the solute-water pair correlation term but showed that the translational component is larger than the orientational one (the former takes 55$\%$−70$\%$ of the total). Later, Kinoshita and coworkers \cite{Kinoshita13,Oshima15,Yoshidome09} examined the translational and orientational components and their pair and many-body correlation terms. It was shown that when the many-body correlation term is also considered, the translational component is much larger than the orientational one. We note, however, that the orientational component is also taken into account in this study using the angle-dependent version \cite{Kinoshita08,Kusalik88a,Kusalik88b,Kinoshita96,Cann97} of the IET.
In Fig. \ref{fig:fig2}(a), upon the burial of a donor and an acceptor in the protein interior, when an IHB is formed (e.g., NH$\cdot$W (Exposed) + O$\cdot$W (Exposed) $longrightarrow$ O$\cdots$HN (Buried); “W” denotes an oxygen atom in a water molecule), we assume that there is no energy change occurring. This assumption is consistent with the results observed by Matubayasi and coworkers \cite{Karino11,Kamo16,Yamamori16} in all-atom MD simulations of a set of protein structures immersed in water: The protein-water electrostatic interaction energy is strongly correlated with the protein intramolecular electrostatic interaction energy; when the former becomes higher, the latter becomes lower (when the former becomes lower, the latter becomes higher); and the magnitudes of changes in the two quantities are not significantly different from each other.
\section{Robustness of the results against uncertainty of the hydrogen-bonding parameters}
\begin{figure}[htpb]
  \centering
    \includegraphics[width=.8\linewidth]{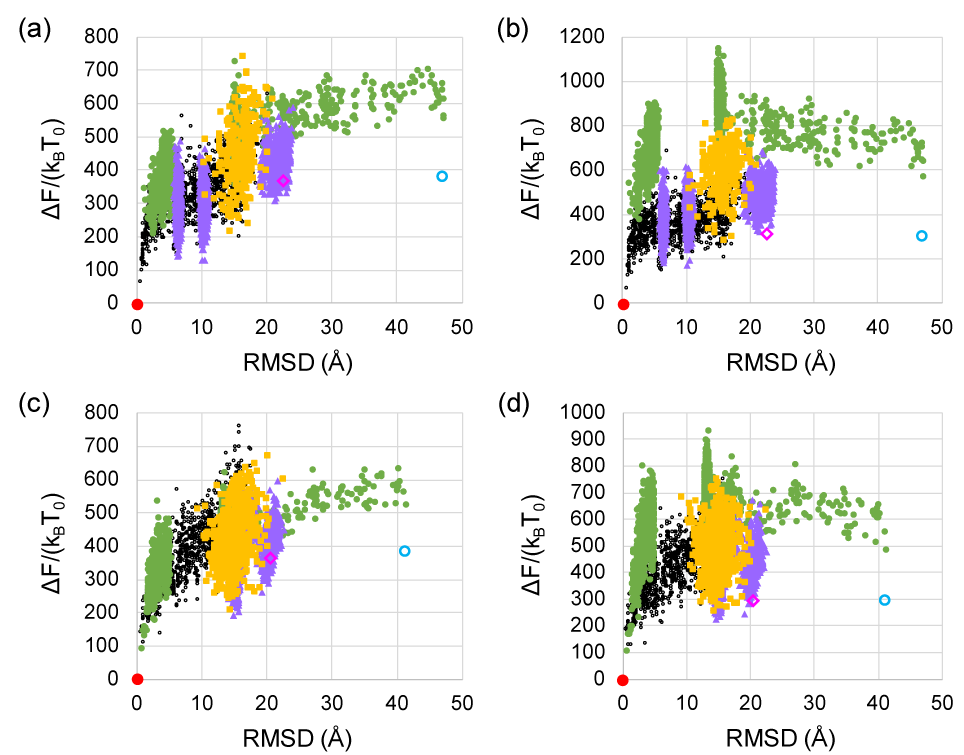}
  \caption{Relation between the RMSD from the native structure in terms of the C$_{\alpha}$ atoms and $\Delta F/(k_BT_0)$. (a), (b) CPB-BD in water. (c), (d) apoPC in water. (a), (c) $7k_BT_0$ and $−14k_BT_0$ in Fig. \ref{fig:fig2}(a) are changed to $5k_BT_0$ and $−10k_BT_0$, respectively. (b), (d) $7k_BT_0$ and $−14k_BT_0$ in Fig. 2(a) are changed to $9k_BT_0$ and $−18k_BT_0$, respectively. Figs. \ref{fig:fig16}(a) and (b) should be compared with Fig.\ref{fig:fig5}(a). Figs. \ref{fig:fig16}(c) and (d) should be compared with Fig. \ref{fig:fig11}(a).
  \label{fig:fig16}}
\end{figure}
Since the results for methanol, ethanol, and cyclohexane are almost the same, they are robust against the uncertainty of the hydrogen-bonding parameters in Figs. \ref{fig:fig2}(b), (c), and (d). The robustness of the result for water was already corroborated in our earlier work (see Fig.\ref{fig:fig2}(a)).12 However, we show it for the two proteins considered in this study, CPB-BD and apoPC. As judged from in Fig. \ref{fig:fig16}, even when $7k_BT_0$ in Fig. \ref{fig:fig2}(a) is changed to $5k_BT_0$ or $9k_BT_0$ ($−14k_BT_0$ is changed to $−10k_BT_0$ or $−18k_BT_0$), the conclusions are not altered at all. Our FEF is capable of discriminating the NS from the decoys as the structure with lowest value of the FEF.
\end{document}